\documentclass[twocolumn,preprintnumbers,amsmath,amssymb,aps,pre,reprint,longbibliography]{revtex4-2}
\usepackage{graphicx}
\usepackage{xcolor}
\begin{document}

\title{Active Microrheology and Dynamic Phases for Pattern Forming Systems with Competing
Interactions
}
\author{
C. Reichhardt and C. J. O. Reichhardt 
} 
\affiliation{
Theoretical Division and Center for Nonlinear Studies,
Los Alamos National Laboratory, Los Alamos, New Mexico 87545, USA
}

\date{\today}

\begin{abstract}
  We consider the driven dynamics of a probe particle moving through an assembly of particles with competing long-range repulsive and short-range attractive interactions, which form crystal, stripe, labyrinth, and bubble states as the ratio of attraction to repulsion is varied. We show that the probe particle exhibits a depinning-like threshold from an elastic regime, where the probe particle is trapped by interactions with the other particles, to a plastic flow regime, where the probe particle can break bonds in the surrounding medium. For a fixed particle density, the depinning threshold and sliding velocity of the probe particle vary nonmonotonically as the attraction term is increased. A velocity minimum appears near the crystal to stripe crossover, and there is a significant increase in the depinning threshold in the bubble regime when the probe particle is strongly confined inside the bubbles. For fixed attractive interaction but increasing particle density, the behavior is also nonmonotonic and there are jumps and drops in the velocity and depinning threshold corresponding to points at which the system transitions between different structures. There are also several distinct flow states that can be characterized by the amount of plastic deformation induced by the probe particle in the surrounding medium. Each flow state generates a different amount of effective drag on the probe particle, and there can be jumps in the velocity-force curve at transitions between the states. We also find that when oriented stripes are present, the probe particle can move along the stripe in an edge transport state that has a finite Hall angle.
\end{abstract}

\maketitle

\section{Introduction}

In active rheology, a probe particle is driven through an
assembly of other particles and produces a response that depends on the
drive amplitude and the nature of the surrounding medium, which may be
solid or fluid-like
\cite{Hastings03,Habdas04,Reichhardt04aa,Squires05,Gazuz09,Khair10,Winter12,Anderson13,Swan13,Benichou13a,Puertas14,Gruber16,Wulfert17,Zia18,Yu20}.
For colloidal assemblies that are
in a glass or crystalline state, there can be a threshold driving
force that must be applied to the probe particle so that it can
transition from
trapped to
flowing 
\cite{Hastings03,Habdas04,Gazuz09,Senbil19,Gruber20,Hopkins22}.
In granular matter, the drag on the probe particle can show a divergence as the
critical jamming density is approached \cite{Drocco05,Candelier10,Kolb13}.
If there is some additional chirality in the system,
the probe particles can exhibit odd viscosity behaviors
such as a Hall effect that can depend on the magnitude of the driving force
\cite{Reichhardt19a,Duclut24}.
In active matter systems, the drag on the probe particle depends on
whether motility induced phase separation
is occurring \cite{Reichhardt15}
and on how deformable the surroundings are \cite{Hopkins22}.

Driving single particles through an assembly of other particles or
over a disordered substrate has also been
employed in hard condensed matter systems.
For example, when dragging individual superconducting vortices
\cite{Straver08,Auslaender09,Reichhardt09a}
or skyrmions \cite{Wang20b,Reichhardt21a},
the probe particle interacts not only
directly with the other particles 
but also with defects or pinning sites in the sample.
In most active rheology systems, such as 
charged colloidal particles, superconducting vortices,
and magnetic skyrmions,
the particle-particle interaction
potential is purely repulsive,
and often consists of a very short range steric repulsion,
as in uncharged colloidal particles and granular matter.
A wide class of particle-based systems with
competing long-range repulsive and short-range attractive
interactions exhibit a variety of pattern forming states as
the relative attraction strength or particle density is varied,
including anisotropic crystals, stripes, labyrinths, bubbles, and void lattices
\cite{Seul95,Stoycheva00,Reichhardt03,Reichhardt04,Mossa04,Sciortino04,Nelissen05,Liu08,Reichhardt10,McDermott14,Liu19,AlHarraq22,Hooshanginejad24}.
Similar patterns can form for systems in which the interactions are
strictly repulsive when multiple length scales are present in the
interaction potential
\cite{Jagla98,Malescio03,Glaser07}.
This type of pattern formation occurs in a variety
of soft matter systems such as colloidal assemblies,
emulsions, binary fluids, and magnetic colloids
with additional capillary interactions
\cite{Malescio03,Glaser07,CostaCampos13}.
Stripe, bubble, and labyrinth patterns also occur in hard condensed
matter systems such as electron liquid crystals
\cite{Fogler96,Moessner96,Cooper99,Fradkin99,Gores07,Zhu09,Friess18},
as well as
several other charge-ordered states where competing interactions
arise \cite{Tranquada95,Reichhardt04a,Mertelj05}.
In multiple component superconductors, the vortex-vortex interactions can have
multiple length scales, giving rise to mesoscale stripe and bubble-like
ordering
in the vortex structures 
\cite{Xu11,Komendova13,Varney13,Sellin13,Brems22}.
Bubbles and stripes can also occur in magnetic systems
such as skyrmion-supporting materials \cite{Reichhardt22a}.

Despite the large number of systems in which
bubble and stripe mesoscale ordering occurs,
active microrheology of bubble and stripe states has not been
studied before now.
Here, we use large-scale numerical simulations 
of a probe particle driven through an assembly of particles
that have a competing short-range attraction
and long-range repulsion.
Previous work with this type of interaction potential has shown that when
the ratio of attraction to repulsion is varied or the density of the system
is changed, the particles can transition among
crystal, anisotropic crystal,
stripe, void lattice, bubble, or
labyrinthine phases
\cite{Reichhardt04,Reichhardt10,Nelissen05,Liu08,McDermott16,Reichhardt24}.
We demonstrate that active rheology of bubble and stripe systems produces a
wide variety of nonmonotonic motion thresholds,
effective viscosities, and
different flow states.
The depinning threshold and type of flow of the probe
particle depend on the patterns formed by the background particles
and the magnitude of the driving force. There is a minimum in the depinning threshold near the crystal-to-stripe
transition that occurs at fixed particle density for increasing attractive interactions.
In the bubble state, there is a sharp increase in the depinning threshold
because the probe particle can remain trapped inside a bubble for an
extended range of drives,
while at higher drives, there is a transition to
a state in which the probe particle hops from bubble to bubble.
We find that systems with competing interactions
generally exhibit a greater variety of distinct flowing phases under
active rheology than what is observed for
purely repulsive interacting particles.
In the stripe phase, 
the probe particle can create large plastic deformations in the stripe,
leading to an increased drag effect;
however, at high drives, the probe particle moves rapidly
enough through the stripe that plastic deformations of the surrounding
medium are reduced or no longer occur,
producing a reduction in the number of fluctuations.
These different dynamic phases are visible as
changes in the velocity force curves and in the
effective drag on the probe particle.
The stripe state may adopt a disordered labyrinth configuration or
an ordered stripe arrangement, depending on the interaction between the
particles. In the ordered stripe state, there can be
transport of the probe particle along the edge of the stripe,
leading to the emergence of a finite Hall angle when the stripes are
not aligned with the driving direction.
At higher drives, the probe particle breaks through the stripe,
leading to increased drag and a reduced Hall angle.

\section{Simulation}
We consider a two-dimensional (2D) system of size $L \times L$, with $L=36$
and periodic boundary conditions
in the $x$- and $y$-directions, containing
$N$ particles.
The system density is $\rho = N/L^2$, and
the particle-particle interaction potential
has both a long-range repulsion and a short-range attraction.
The probe particle is subjected to
an external driving force $F_{D}$
applied along the positive $x$-direction.
The particle dynamics are obtained by integrating
the following overdamped equation of motion:
\begin{equation}
\eta \frac{d {\bf R}_{i}}{dt} =
-\sum^{N}_{j \neq i} \nabla V(R_{ij}) +
        {\bf F}_{D}^i \ 
\end{equation}
where $R_{ij}=|{\bf R}_i-{\bf R}_j|$ is the distance between particles
$i$ and $j$.
For the non-driven particles, ${\bf F}_D^i=0$.
The damping term is set to $\eta=1.0$.
The particle-particle interaction forces in the first term on the right
hand side are obtained from the following potential:
\begin{equation}
V(R_{ij}) = \frac{1}{R_{ij}} - B\exp(-\kappa R_{ij}) \ .
\end{equation}
The first term is the repulsive Coulomb interaction,
and the second term is an attraction of strength
$B$ and inverse range $\kappa$.
In this work we fix $\kappa=1.0$ while varying $B$ and the density $\rho$;
however, we note that the type of patterns that form also depend on $\kappa$.
The particle-particle interaction is repulsive at large distances and
attractive at intermediate distances.
At very short range, the Coulomb term is dominant, so the particles cannot
all collapse onto a point.
To treat the $1/r$ interaction potential, we
employ the real space Lekner summation technique,
as used in previous work \cite{Reichhardt03,McDermott14}.
The interaction potential in Eq.~(2) has previously been
shown to produce
crystal, stripe, bubble, and void lattices
as the attraction strength $B$ is varied
\cite{Reichhardt03,Reichhardt03a,Reichhardt04,Reichhardt10,McDermott14}.

The initial particle configurations are
obtained by placing the particles in a uniform triangular lattice
and allowing them to relax into the patterned states.
We have also performed simulated annealing
by starting from a high
temperature and slowly cooling the system to $T = 0.0$.
We obtain similar patterns and dynamics for either initialization method.
After the system has been initialized,
we apply a driving force to the probe particle
which we increase in increments of 
$\delta F_{D} = 0.01$.
We spend 6200 to 10000 simulation time steps at each drive increment
before increasing the drive again.
We find the time-averaged velocity
$\langle V\rangle=\langle v_i \cdot {\bf \hat x}\rangle$ and
$\langle V_y\rangle=\langle v_i \cdot {\bf \hat y}\rangle$ of the probe
particle both parallel and perpendicular to the driving direction.

\begin{figure}
\includegraphics[width=\columnwidth]{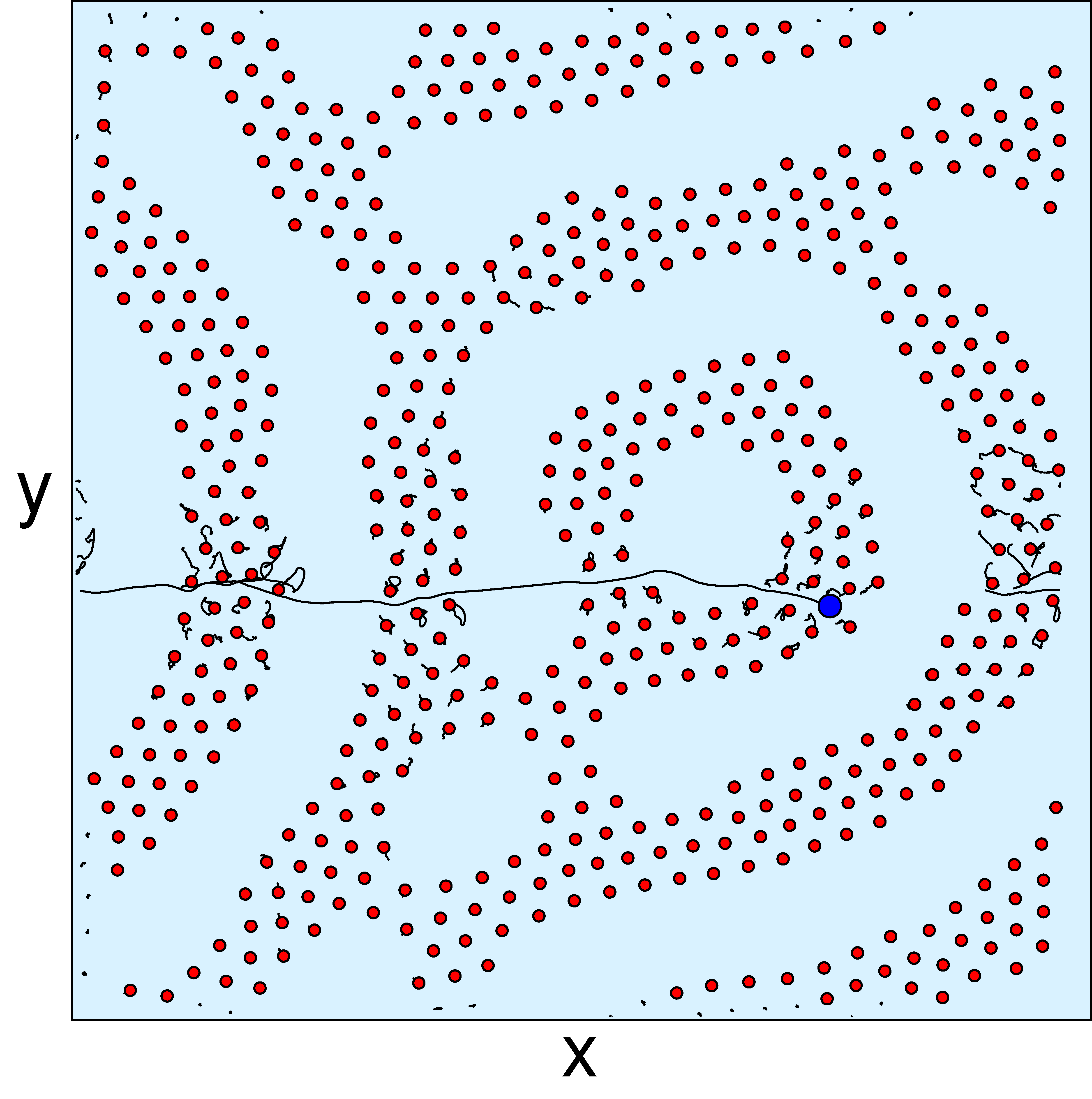}
\caption{Locations (dots) and trajectories (lines) of the probe
particle (blue) and surrounding particles (red) in a system with  
$\rho = 0.44$ and $B = 2.0$ which forms a disordered stripe or labyrinthine
phase. At $F_{D} = 0.7$, the probe particle can break through
the stripe structure. In all results presented in this work, the probe particle
and the background particles have the same size and the same charge, but
the probe particle has been drawn larger than the background particles
for clarity.
} 
\label{fig:1}
\end{figure}

\section{Results}

\begin{figure}
\includegraphics[width=\columnwidth]{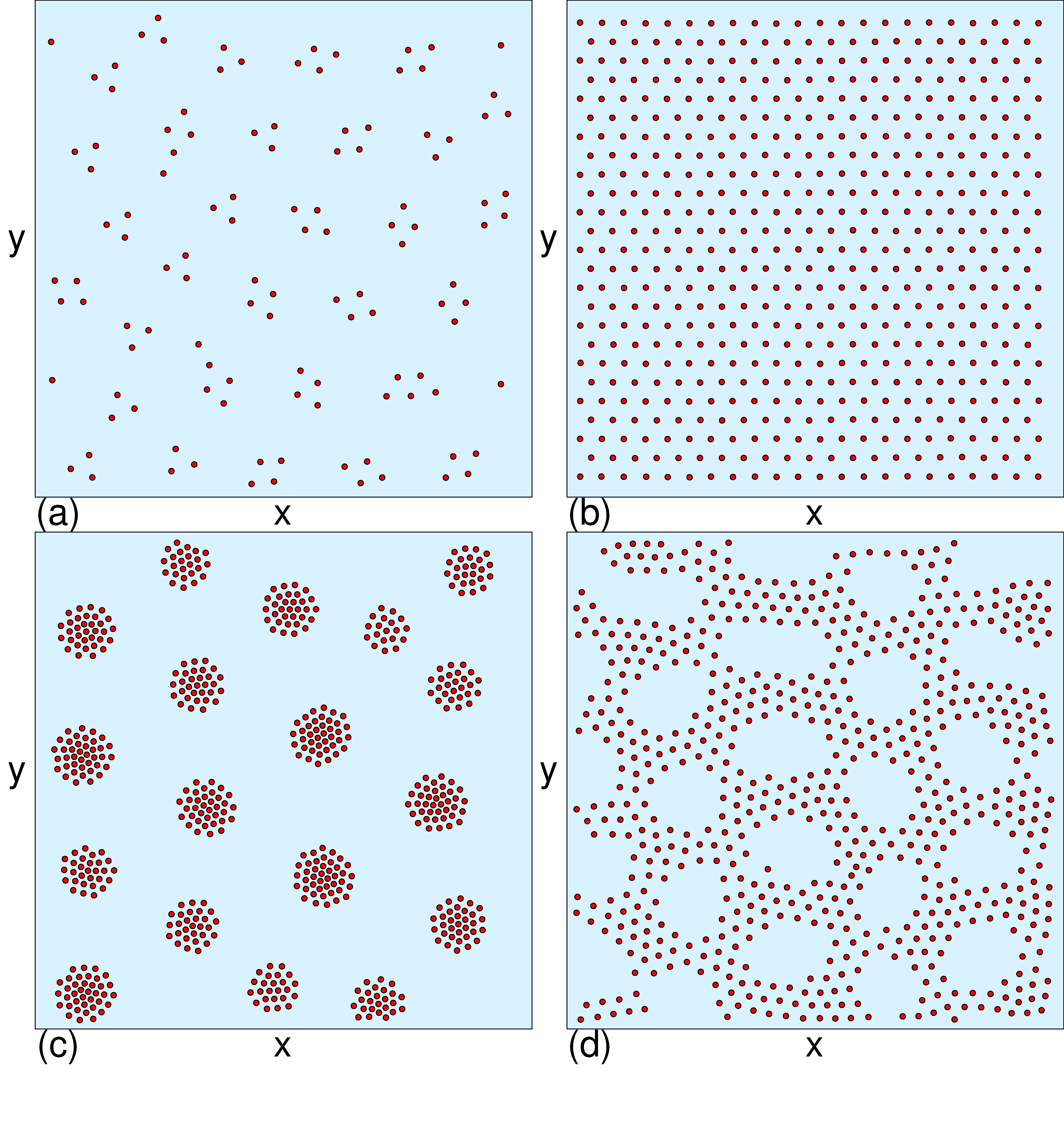}
\caption{Images of particle locations for
representative examples of the different patterns that can form
in the system for varied parameters. Here $F_D=0$.
(a) A low density bubble phase
at $\rho = 0.0952$ and $B =1.8$.
(b) A uniform crystal at $B = 1.6$ and
$\rho = 0.44$.
(c) A bubble state at $\rho= 0.44$ and $B = 2.4$.
(d) A void lattice at $\rho = 0.52$ and $B  =2.0$.
}
\label{fig:2}
\end{figure}

In Fig.~\ref{fig:1}, we show the locations and trajectories of the probe
particle and the surrounding particles
for a system in the disordered stripe regime
at $\rho = 0.44$ and $B = 2.0$.
For $F_D=0.7$, as shown in the figure, the probe particle can break through
the stripe structure and cause distortions within it.
Figure~\ref{fig:2} illustrates representative examples of the different types
of patterns that form as $B$ and $\rho$ are varied.
There is a low density bubble phase for $\rho=0.0952$ and $B=1.8$ in
Fig.~\ref{fig:2}(a), a crystal phase at $\rho=0.44$ and $B=1.6$ in
Fig.~\ref{fig:2}(b),
a bubble state for $\rho =0.44$ and $B = 2.3$ in Fig.~\ref{fig:2}(c),
and a void lattice at $\rho = 0.52$ and $B =2.0$ in Fig.~\ref{fig:2}(d).

\begin{figure}
\includegraphics[width=\columnwidth]{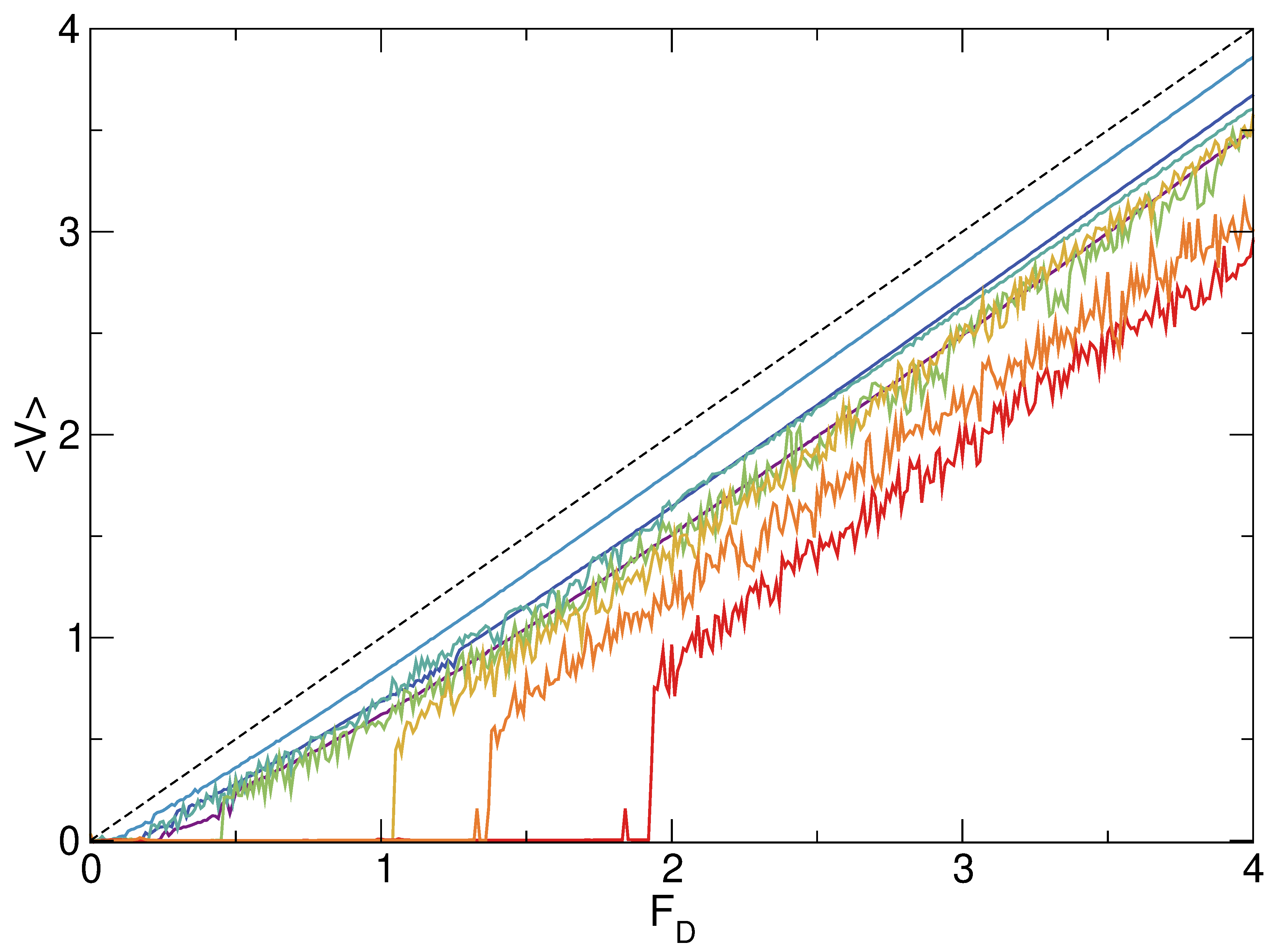}
\caption{The velocity-force curves $\langle V\rangle$ vs $F_D$
for the probe particle
for the system from Fig.~\ref{fig:1} with $\rho = 0.44$
in the crystal state at
$B = 0.2$ (violet), $B=0.8$ (dark blue),
and $B=1.9$ (light blue);
the stripe state at
$B=2.0$ (teal);
and the bubble state at
$B=2.2$ (green),
$B=2.4$ (yellow),
$B=2.5$ (orange),
and $B=2.6$ (red).
The lowest threshold and highest sliding velocity occur for $B  = 1.9$.
In the bubble phase, the depinning threshold increases with increasing $B$.
}
\label{fig:3}
\end{figure}

In Fig.~\ref{fig:3}, we plot some representative velocity-force
curves for the probe particle from the system in Fig.~\ref{fig:1} at
a fixed density of $\rho = 0.44$
for the crystal state at $B = 0.2$, $B=0.8$, and $B=1.9$, the stripe
state at
$B=2.0$, 
and the bubble state at $B=2.2$, $B=2.4$, $B=2.5$, and $B=2.6$.
The black dashed line indicates
the expected velocity-force curve for a freely moving particle.
Each curve shows a finite threshold for motion followed by
a sliding phase.
The lowest depinning threshold occurs
in the crystal regime at $B = 1.9$.
The threshold increases rapidly upon entering the bubble state and
continues to increase with increasing $B$. There are strong
velocity fluctuations
in the sliding phase when the particle is jumping from bubble to bubble.
The velocity response is nonmonotonic in the sliding regime, as
exemplified by the fact that at
$F_{D} = 1.0$, the velocity is low for $B = 0.2$, gradually increases
with increasing $B$ until $B = 1.9$, and then decreases again as
$B$ increases further.

\begin{figure}
\includegraphics[width=\columnwidth]{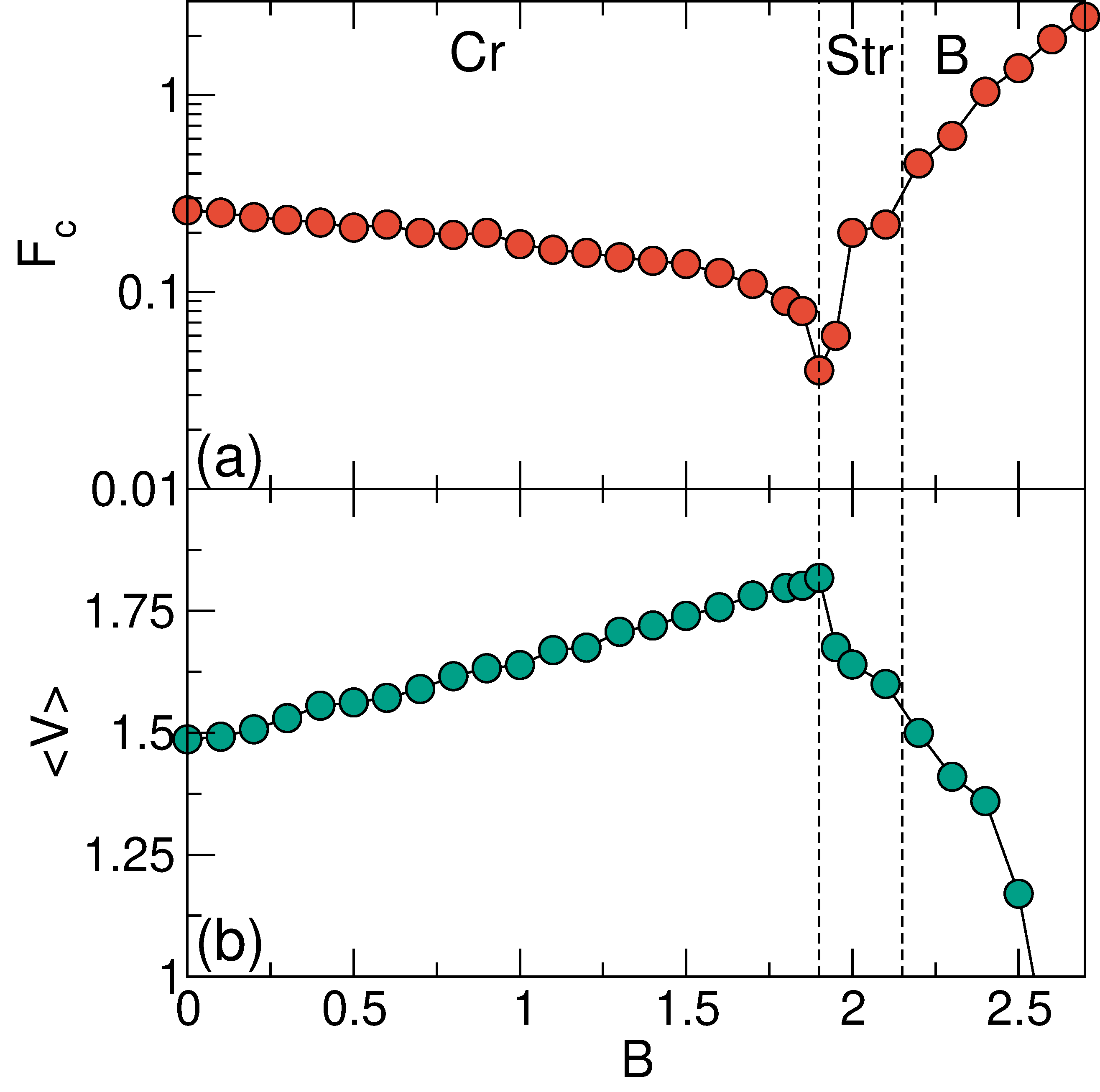}
\caption{
(a) The depinning threshold $F_{c}$ vs $B$ for the system in
Fig.~\ref{fig:3} at $\rho = 0.44$.
(b) The corresponding velocity $\langle V\rangle$ at
$F_{D}= 1.0$ vs $B$. The dashed lines separate the crystal (Cr), 
stripe (Str), and bubble (B) phases.
The threshold $F_c$ passes through
a minimum near the crystal to stripe transition
and rapidly increases with increasing $B$
in the bubble phase.
}
\label{fig:4}
\end{figure}

In Fig.~\ref{fig:4}(a), we plot the threshold force $F_{c}$ for motion
versus $B$ at fixed $\rho = 0.44$,
and in Fig.~\ref{fig:4}(b) we show the
corresponding value of $\langle V\rangle$ at $F_{D} = 1.0$.
The dashed lines indicate the boundaries of the crystal (Cr),
stripe (Str), and bubble (B) states.
We find that $F_{c}$ decreases with increasing $B$ in the crystal
state, and that at the same time there is a
linear increase in $\langle V\rangle$.
A dip in $F_{c}$ appears at $B = 1.9$ just before the crystal-to-stripe
transition, and is associated with a peak in
$\langle V\rangle$.
In the stripe phase, $F_{c}$ increases with increasing $B$ while
$\langle V\rangle$ drops, and there is a more rapid increase of
$F_c$ with increasing $B$ in the bubble phase.

\begin{figure}
\includegraphics[width=\columnwidth]{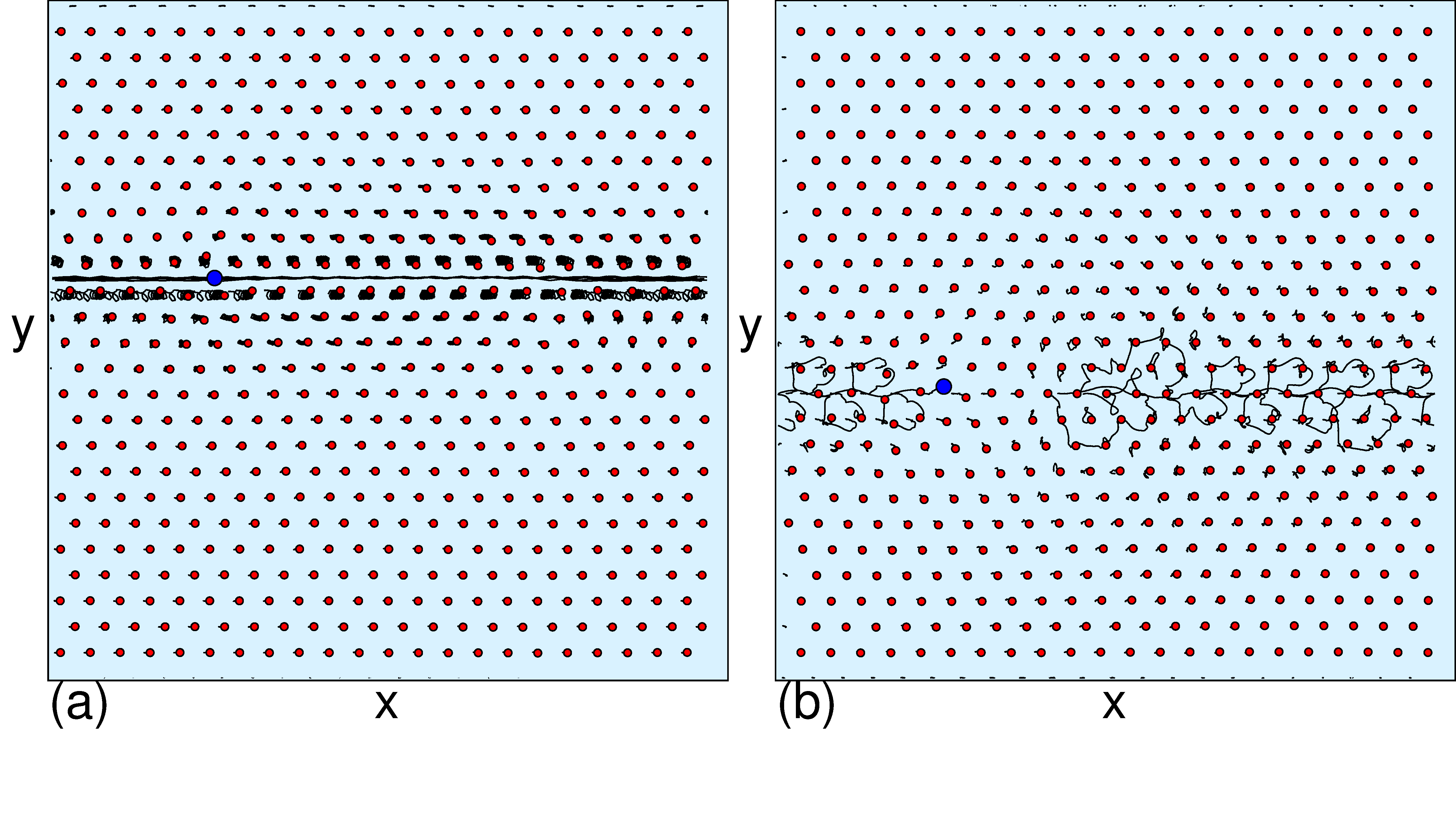}
\caption{Locations (dots) and trajectories (lines) of the
probe particle (blue) and surrounding particles (red) in the
crystal phase at
$\rho = 0.44$ and $B=0.2$.
(a) At $F_{D} = 1.0$, the probe particle passes between two rows
of the background lattice without inducing plastic distortions.
(b) At $F_{D} = 0.3$, just above the threshold $F_c$,
the motion of the probe particle causes
exchanges of the surrounding particles.
}
\label{fig:5}
\end{figure}

In the crystal regime, the probe particle depins and flows
along a one-dimensional path between rows of the background particles,
as shown in Fig.~\ref{fig:5}(a) at $B = 0.2$  and $F_{D}  = 1.0$.
At lower drives, this channel motion is unstable and the probe particle
moves by exchanging places with the background particles, resulting in
the appearance of
plastic deformations in the surrounding lattice, as illustrated in
Fig.~\ref{fig:5}(b) at $F_D=0.3$, just above the threshold drive.
Since the plastic deformations result in an increased drag on the
probe particle, there is a two-step feature in the velocity-force
curve in Fig.~\ref{fig:3}, with a transition from a low drive
motion with strong plastic distortions to an ordered flow at high drives.

For the stripe forming states, the probe particle is initially pinned since
there are no rows providing easy-flow channels for motion. In order to
translate, the probe particle must break out of the caging potential
produced by the surrounding particles and force its way across the
stripe.
This gives a larger value of $F_{c}$
compared to the crystal regime and
also increases the effective drag on the
probe particle.
In the bubble phase, the probe particle becomes trapped by the
attractive portion of the interparticle
interaction potential.
As $B$ increases, the attractive portion of the potential becomes stronger,
increasing the barrier that must be overcome before the probe particle
can jump from bubble to bubble.
The dip in $F_{c}$  near the crystal to stripe
transition in Fig.~\ref{fig:4}(a) occurs
when the attractive and repulsive forces on the probe particle are
exactly balanced.
Within the crystal phase, the probe particle is trapped by the
caging effect from the repulsion of the other particles.
For $B = 0.0$, this repulsive barrier is maximum.
As $B$ increases, the net repulsion is reduced since it is
counterbalanced by
the attractive force.
The repulsion reaches its minimum
at the onset of the stripe phase when the attractive part of the
interaction force begins to dominate the caging of each particle, pulling
the particles together into stripe structures. Like the background
particles, the probe particle also becomes trapped by
the attractive potion of the
interaction potential.
The magnitude of the attractive force increases
when the stripes transform into bubbles,
and as a result the threshold force $F_c$ also increases at the
stripe-to-bubble transition.

\begin{figure}
\includegraphics[width=\columnwidth]{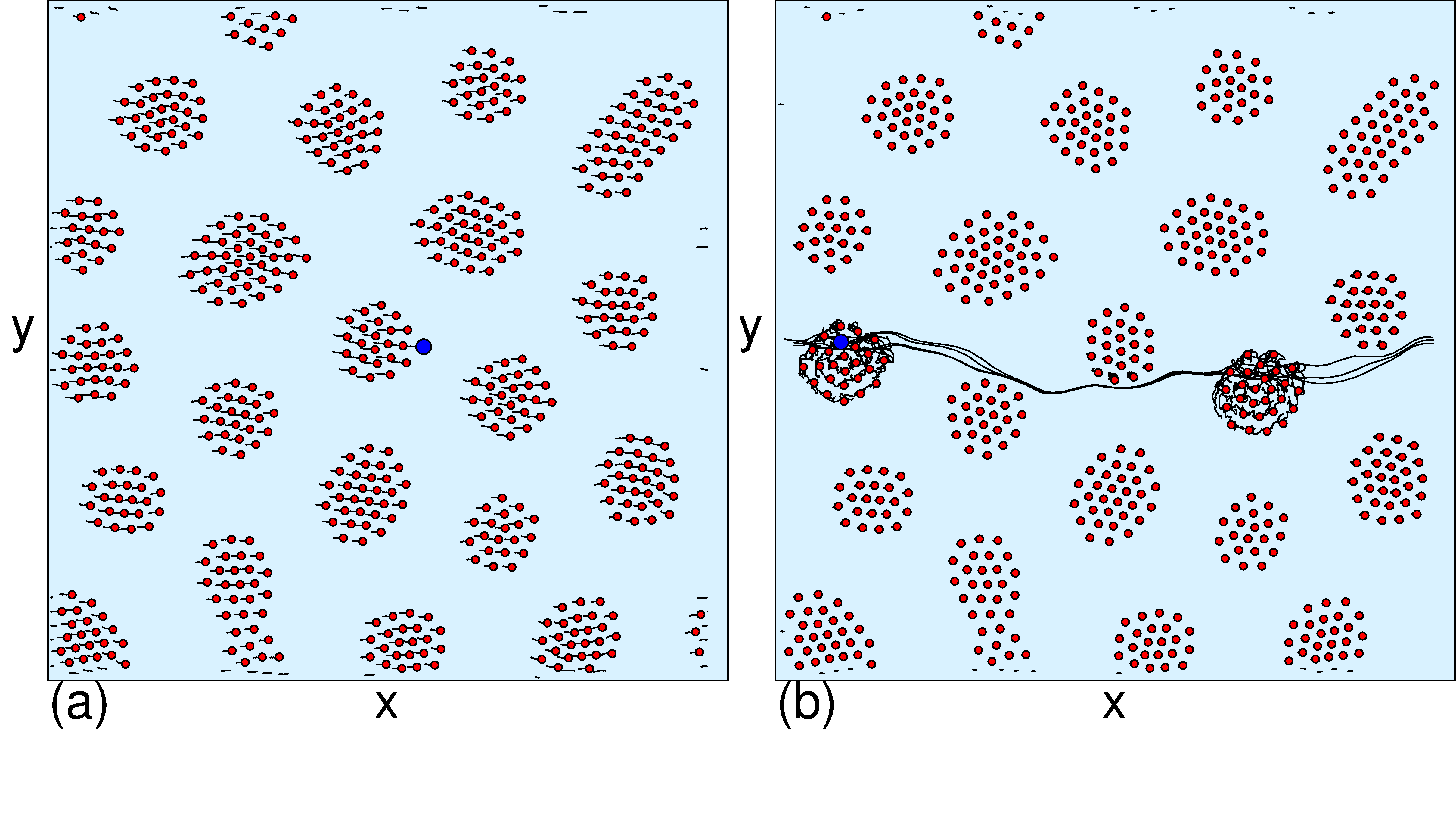}
\caption{Locations (dots) and trajectories (lines) of the probe
particle (blue) and surrounding particles (red) in the bubble phase  
at $\rho = 0.44$ and $B = 2.4$.
(a) At $F_{D} = 0.5$, the probe particle remains trapped in a
bubble, but the entire system translates
as the bubble is dragged by the probe particle. 
(b) At $F_{D} = 1.5$, the probe
particle jumps from bubble to bubble,
creating plastic distortions in the bubbles.  
}
\label{fig:6}
\end{figure}

The trapped to flowing transition of the probe particle can also be
viewed as
an elastic to plastic transition.
In Fig.~\ref{fig:6} we show the trajectories of the particles
for the system from Fig.~\ref{fig:4} at $\rho = 0.44$ and $B = 2.4$ in the
bubble phase.
At $F_D=0.5$ in Fig.~\ref{fig:6}(a),
the probe particle remains trapped in a bubble,
but the entire system translates 
as this bubble is dragged.
As indicated in the figure, the probe particle works its way to the front
of the bubble that it is dragging.
Here, the velocity of the probe particle is the same as the
velocity of the background particles.
The velocity in this elastic
flow state decreases as $1/N$, where $N$ is the number of particles in
the system, so for large systems,
the elastic velocity becomes 
very low and the system can be
regarded as effectively pinned.
In Fig.~\ref{fig:6}(b) at $F_{D} = 1.5$, the probe
particle jumps from bubble to bubble,
creating plastic distortions as it moves through the bubble
lattice.

\begin{figure}
  \includegraphics[width=\columnwidth]{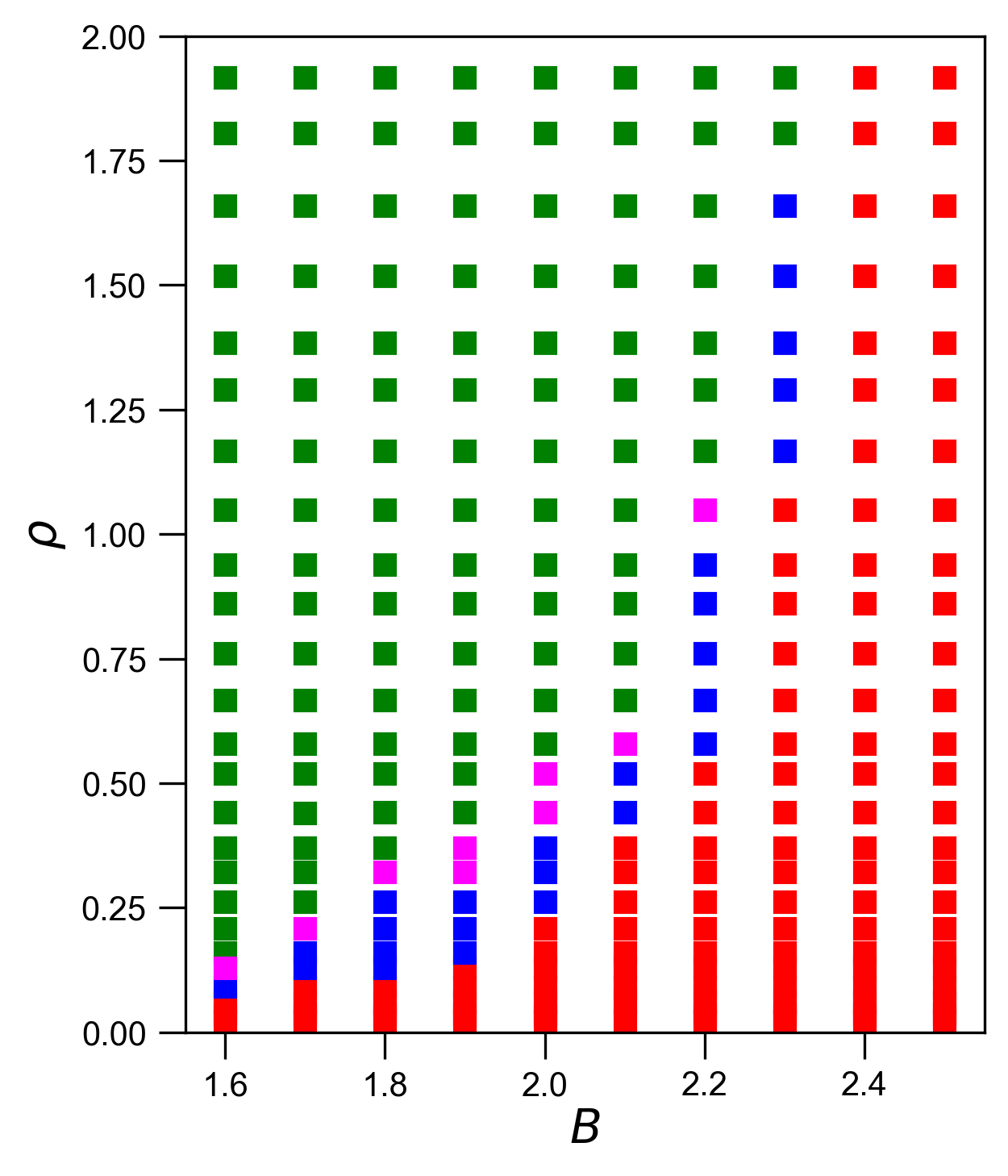}
\caption{Phase diagram as a function of
$\rho$ vs $B$ in the absence of a driven probe particle
highlighting where the system forms
a triangular lattice (green), void lattice (magenta), stripes
(blue), and bubbles (red).}
\label{fig:7}
\end{figure}

\begin{figure}
\includegraphics[width=\columnwidth]{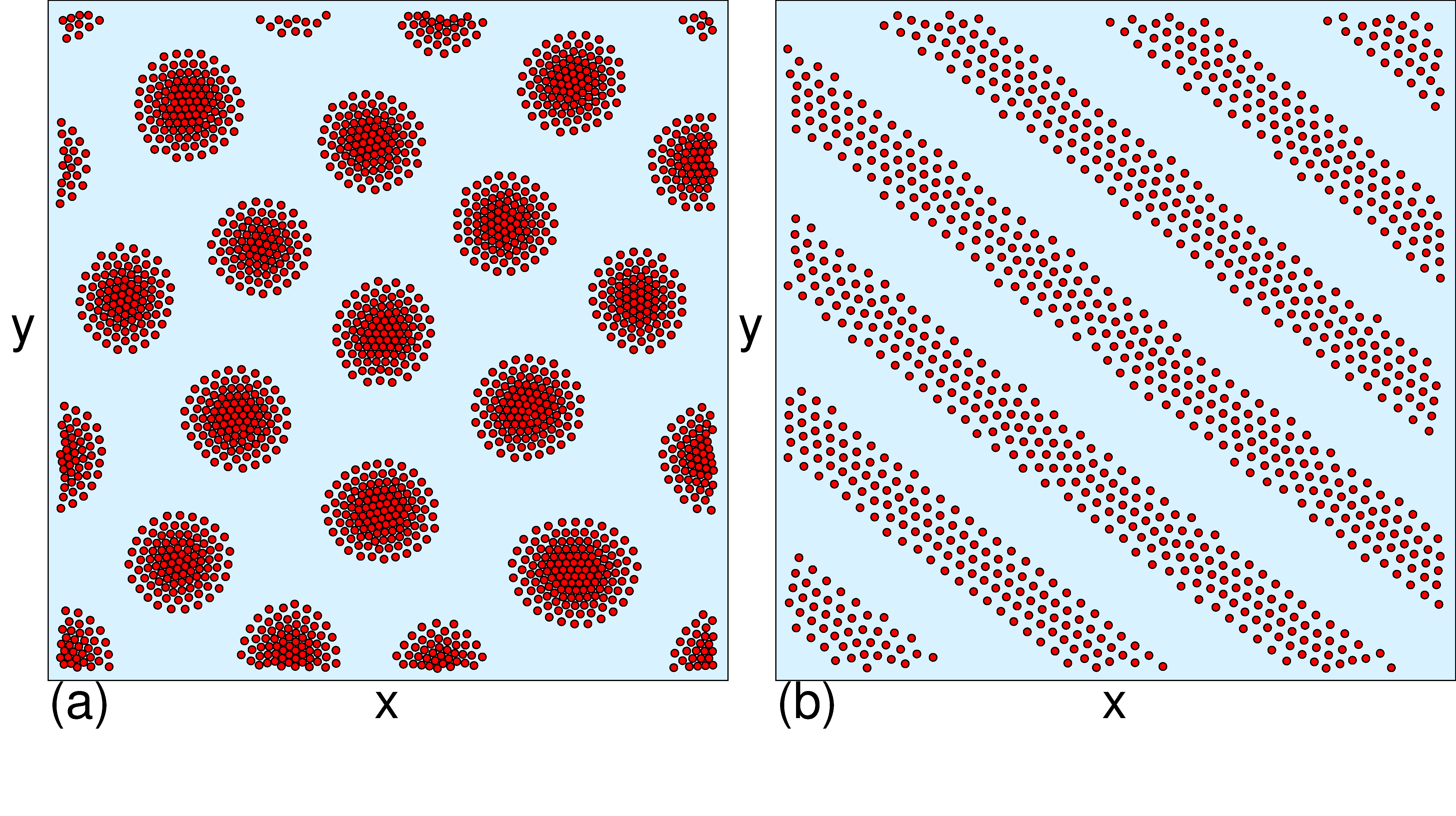}
\caption{Image of particle positions in the absence of a driven
probe particle.  
(a) A bubble lattice at $\rho = 1.8$ and $B = 2.4$, where
there is a density gradient inside the individual bubbles.
(b) An oriented stripe state
at $B = 2.2$ and $\rho = 0.86$.
}
\label{fig:8}
\end{figure}

In Fig.~\ref{fig:7}, we plot a phase diagram
as a function of $\rho$ versus $B$
highlighting where the system, in the absence of a probe
particle, forms a triangular
lattice, void lattice, stripes, and bubbles.
Generally, for a fixed value of $B$, clumps or bubbles appear when the
density is low.
The size and shape of the clump
depend on the value of $B$ and $\rho$.
The clumps can be small or even disordered,
as illustrated in Fig.~\ref{fig:2}(a),
or large and form a lattice, as shown in
Fig.~\ref{fig:8}(a) at $\rho = 1.6$
and $B = 2.4$.
There is also a density gradient within the large individual bubbles.
The stripes can form disordered states, which is typical when the stripes
are thin, but they can also
form ordered structures at higher densities,
as shown in Fig.~\ref{fig:8}(b) at $B = 2.2$ and $\rho = 0.86$.
In general, many of the stripe and bubble states exhibit
aspects of metastability,
so the ability to form ordered stripes versus disordered stripes
in a real experiment would be controlled by the boundary conditions and
the manner in which the system is prepared.
The ordered stripe phase we illustrate in Fig.~\ref{fig:8}(b)
could break into domains of different orientation
if we considered a system of much larger size;
however, we can still examine the active rheology for
both ordered and disordered states
since it should be possible to prepare fully ordered stripe systems
in an experiment.
The phase diagram in Fig.~\ref{fig:7} indicates that as $B$ increases,
stripes and voids do not form until the density $\rho$ is higher.

\begin{figure}
\includegraphics[width=\columnwidth]{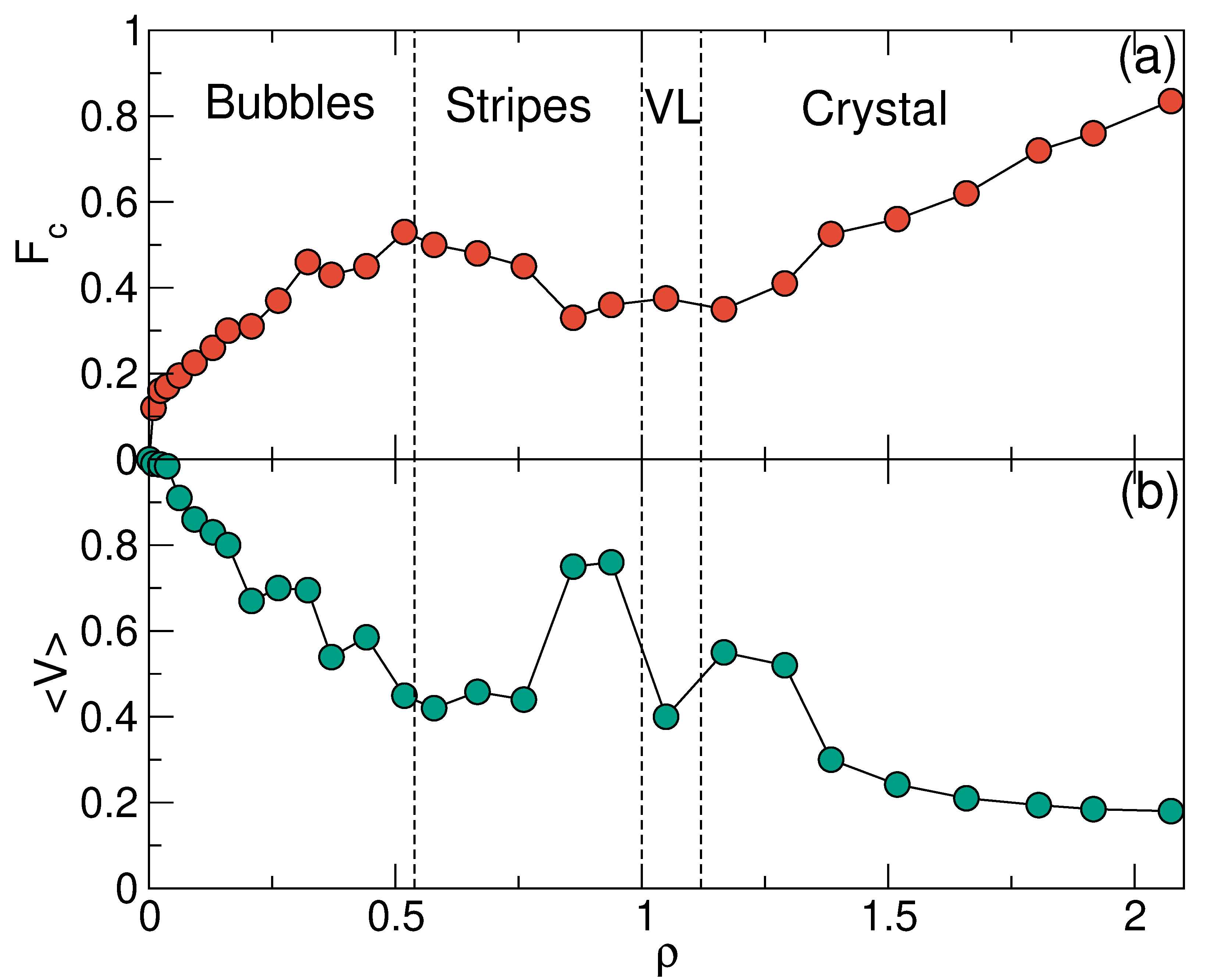}
\caption{(a) The depinning threshold $F_{c}$ vs $\rho$ for
a system with $B = 2.2$.
(b) The corresponding velocity $\langle V\rangle$ at $F_{D}= 1.0$
vs $\rho$.
The dashed lines separate the bubbles, stripes, void lattice (VL),
and crystal phases.
}
\label{fig:9}
\end{figure}

In Fig.~\ref{fig:9}(a), we plot the depinning threshold $F_{c}$
versus $\rho$ at a fixed $B = 2.2$,
while in Fig.~\ref{fig:9}(b) we show
the corresponding $\langle V\rangle$ at a drive of $F_D=1.0$
versus $\rho$.
The dashed lines show where transitions among
the bubble, stripe, void lattice, and crystal
states occur.
At low $\rho$ where the system forms small bubbles,
$F_{c}$ has a low value.
As $\rho$ increases, the bubbles grow in size,
and the threshold $F_c$ also increases until it reaches
a local maximum at the bubble to stripe crossover near $\rho = 0.525$.
The velocity drops to a local minimum in the disordered stripe phase 
that spans $0.525 < \rho < 0.75$.
An oriented stripe phase
appears for $0.75 \leq \rho < 1.0$ that is associated
with a local maximum of the velocity and a dip in the depinning threshold.
When the stripes are oriented, the probe particle is able to glide along
the edge of a stripe, which we discuss in further detail below.
The velocity is reduced
in the void lattice state since the probe particle
must break the particle-particle interaction bonds
in order to move through the structure.
In the crystal state, $F_c$ grows with increasing $\rho$ since the barrier
to motion for high particle densities arises from the
shorter range portion of the repulsive term in the interaction potential,
and this barrier increases as the particles are pushed closer together.
There is a jump up in velocity at the transition to the crystal state
because the probe particle is able to establish a one-dimensional channel
flow between the rows of the background lattice.
For $\rho > 1.6$, the depinning transition becomes plastic and the velocity
at the depinning transition drops due to the increased
amount of plastic deformation;
however, for drives greater than $F_D=1.0$,
there is a jump up in the velocity when the probe particle is able to
transition to interstitial flow that does not induce
plastic deformation in the surrounding medium.
For systems with purely repulsive interactions, the depinning
threshold increases monotonically and the velocity at fixed drive
decreases monotonically
with increasing $\rho$. For our competing interaction system,
both the depinning threshold and the velocity
at fixed drive change strongly nonmonotonically with
increasing $\rho$.

\begin{figure}
\includegraphics[width=\columnwidth]{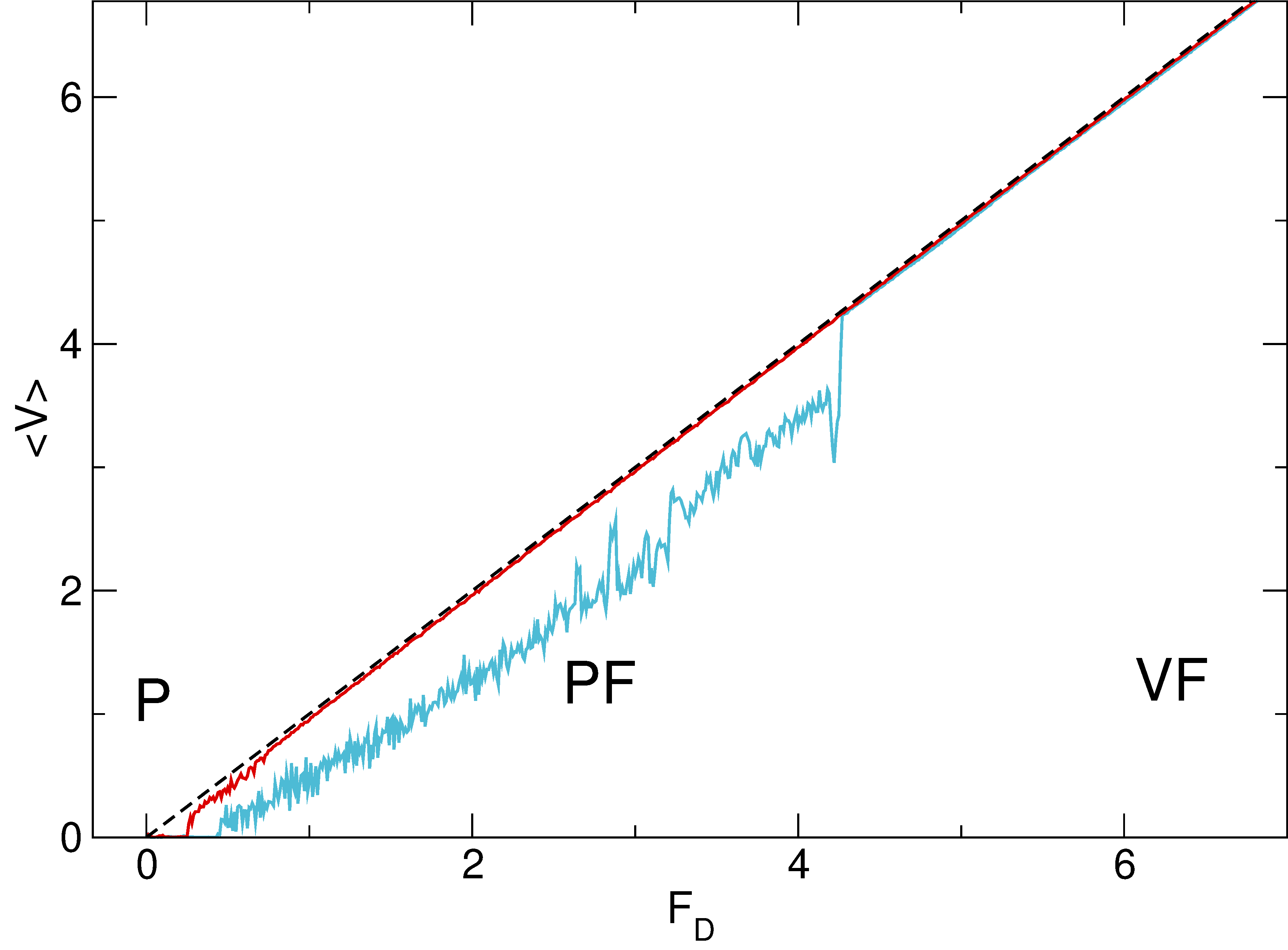}
\caption{$\langle V\rangle$ vs $F_{D}$ in the bubble state at $B=2.2$
for $\rho=0.52$ (blue) and $\rho=0.129$ (red).
The pinned (P) state, plastic flow (PF) state, and viscous
flow (VF) states are labeled for the $\rho=0.52$ system.
The transition to viscous flow occurs at a much lower
drive for the $\rho=0.129$ system where the bubbles are very small.
The dashed line is the expected velocity-force curve for a free particle.
}
\label{fig:10}
\end{figure}

We find distinctive flow regimes in the bubble system.
In Fig.~\ref{fig:10}, we plot $\langle V\rangle$ versus $F_{D}$ for the
bubble state at $B=2.2$ for densities of
$\rho = 0.52$ and $\rho=0.129$, along with a dashed line indicating the
expected response for a free particle.
When $\rho=0.52$,
there is a pinned regime
followed by a plastic flow (PF) regime with large velocity
fluctuations and significant distortions in the surrounding media.
At higher drives, the velocity jumps up nearly to the free flow value and
the velocity fluctuations are strongly reduced.
In this state, the probe particle is moving fast enough that the background
particles cannot respond 
rapidly enough for plastic distortions to occur.
We call this the viscous flow (VF) regime
because the probe particle is still moving more slowly than
a free particle since
the interactions with the background particles enhance the
effective viscosity.
The onset of the viscous flow regime depends strongly on the type of
pattern that forms.
Generally, the PF-VF transition falls at low drive values for
small $\rho$, and reaches its maximum value
in the stripe regime.
In Fig.~\ref{fig:10}, the PF-VF transition for $\rho=0.129$, where the bubbles
are small, occurs at a much lower value of $F_D$ than for the $\rho=0.52$
system.
When the bubbles are small, the amount of possible plastic deformation that
can be produced by the moving probe particle is reduced, so a lower threshold
drive is needed to eliminate the plastic deformations and produce the
viscous flow state.

\begin{figure}
\includegraphics[width=\columnwidth]{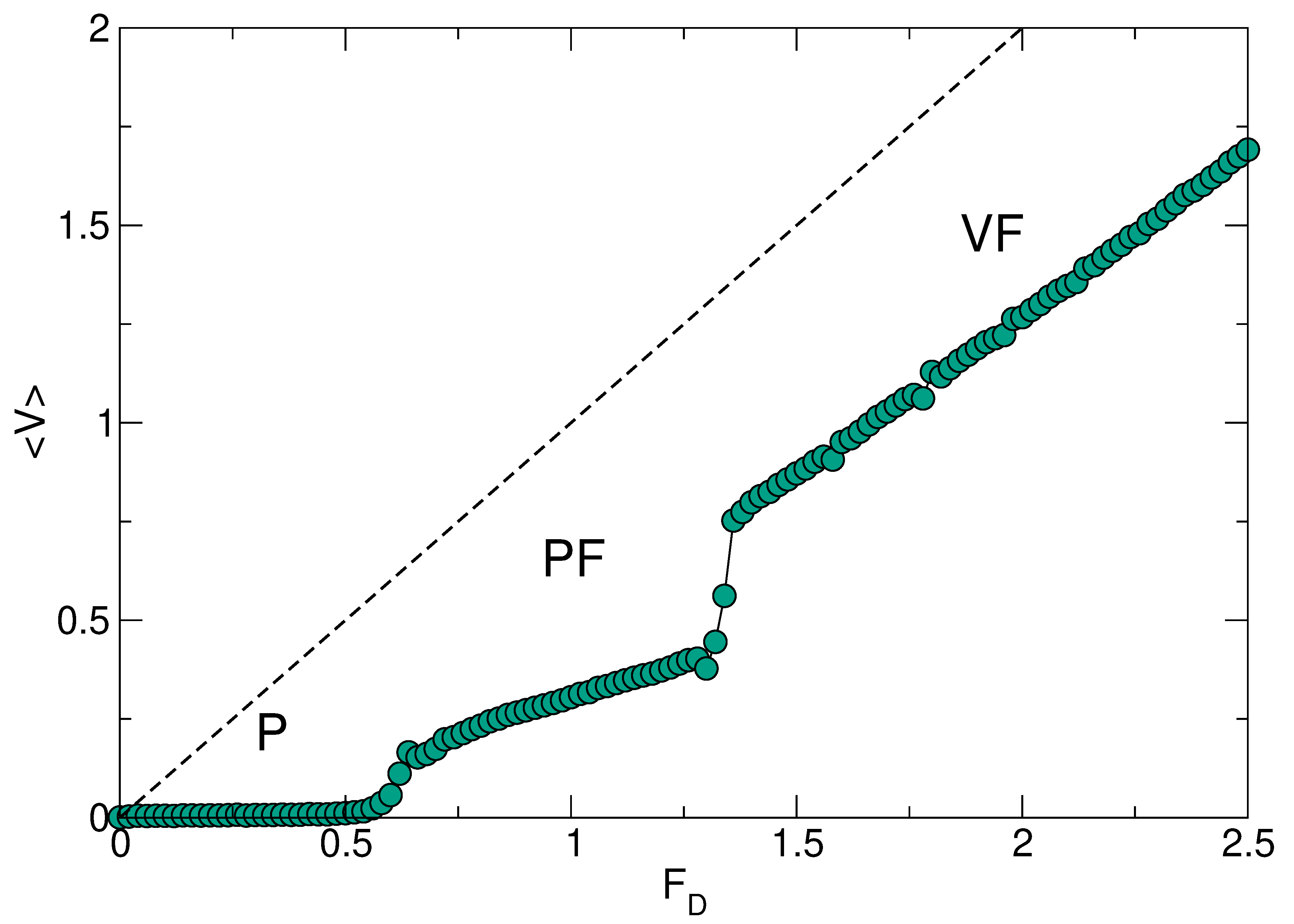}
\caption{$\langle V\rangle$ vs $F_{D}$ for a uniform crystal
state at $\rho = 1.38$ and $B = 2.2$
showing a pinned (P) phase, a plastic flow (PF) phase,
and a viscous flow (VF) phase. The dashed line is the
expected velocity response
for a free particle.
}
\label{fig:11}
\end{figure}

In the uniform crystal phase, there can also be a plastic flow (PF) regime
just above depinning, as shown in Fig.~\ref{fig:11} where we plot
$\langle V\rangle$ versus $F_D$ for a crystal state at
$B=2.2$ and $\rho=1.38$.
At higher drives,
a transition to
a VF regime occurs when the probe particle begins to move between the
rows of the background particles.
The velocity response is reduced in the PF regime due to the plastic
distortions, and there is a jump up in the velocity at the transition
to the VF regime.
As $\rho$ increases, the velocity in the VF regime decreases because
the probe particle interacts with a greater number of background particles,
resulting in a higher effective viscosity.
Additionally, the velocity fluctuations in the PF regime are smaller for
the uniform crystal state than in the bubble or stripe states since the
plastic motion occurs in a more correlated fashion in the uniform
crystal.

\begin{figure}
\includegraphics[width=\columnwidth]{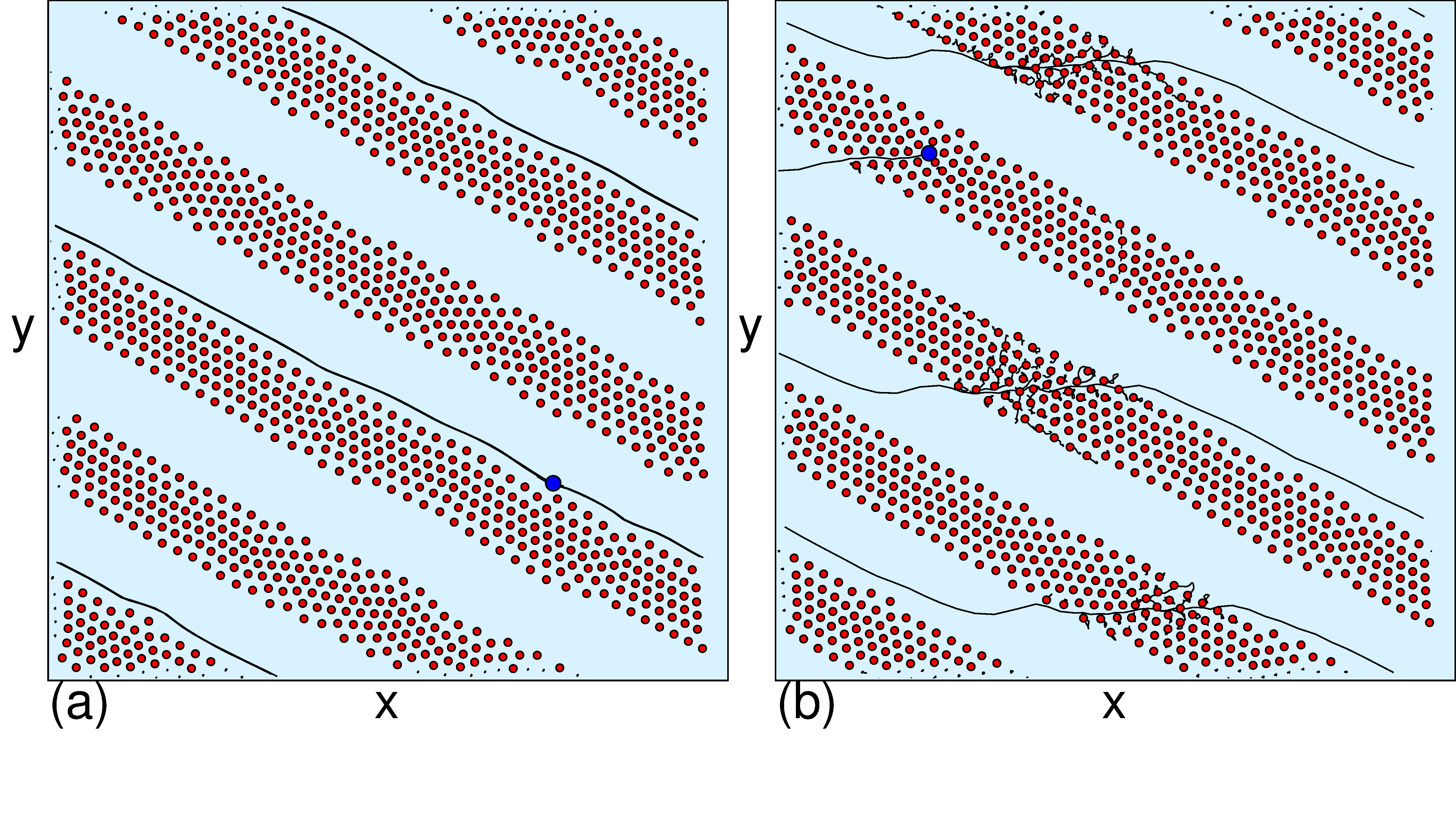}
\caption{Locations (dots) and trajectories (lines) of the probe
particle (blue) and surrounding particles (red) in the stripe state
at $B=2.2$ and $\rho=0.94$.
(a) At $F_D=0.5$, the probe particle moves along the front edge
of the stripe and does not follow the driving direction.
(b) $F_D=1.05$, just above the drive where the probe particle is able to
break through the stripe and create plastic deformations.
}
\label{fig:12}
\end{figure}

Figure~\ref{fig:9} shows that as $\rho$ increases, two regimes of
stripe states appear:
disordered stripes for $0.525 < \rho <  0.75$,
and oriented stripes for $0.75 \leq \rho < 1.0$.
The initial depinning in the ordered stripe state occurs when the probe
particle is able to escape from inside the stripe but cannot reenter
the stripe, so that the particle runs along the edge of the stripe
as illustrated in Fig.~\ref{fig:12}(a) for a system with $B=2.2$ and
$\rho=0.94$ 
at $F_D=0.5$.
In this case, the probe particle is moving along the front edge of the
stripe since the attractive portion of the potentials of the particles inside
the stripe has captured the probe particle.
This edge motion occurs without plastic deformations, and produces 
the large velocity in the stripe phase
at fixed $F_{D} = 1.0$ that is shown in
Fig.~\ref{fig:9}(b). Since there is no mechanism that would cause the stripes
to order along the $x$ direction, when the probe particle follows the edge of
the stripe it does not move along the driving direction but instead travels
at an angle to the driving direction, resulting in the emergence of a
finite Hall angle.

\begin{figure}
\includegraphics[width=\columnwidth]{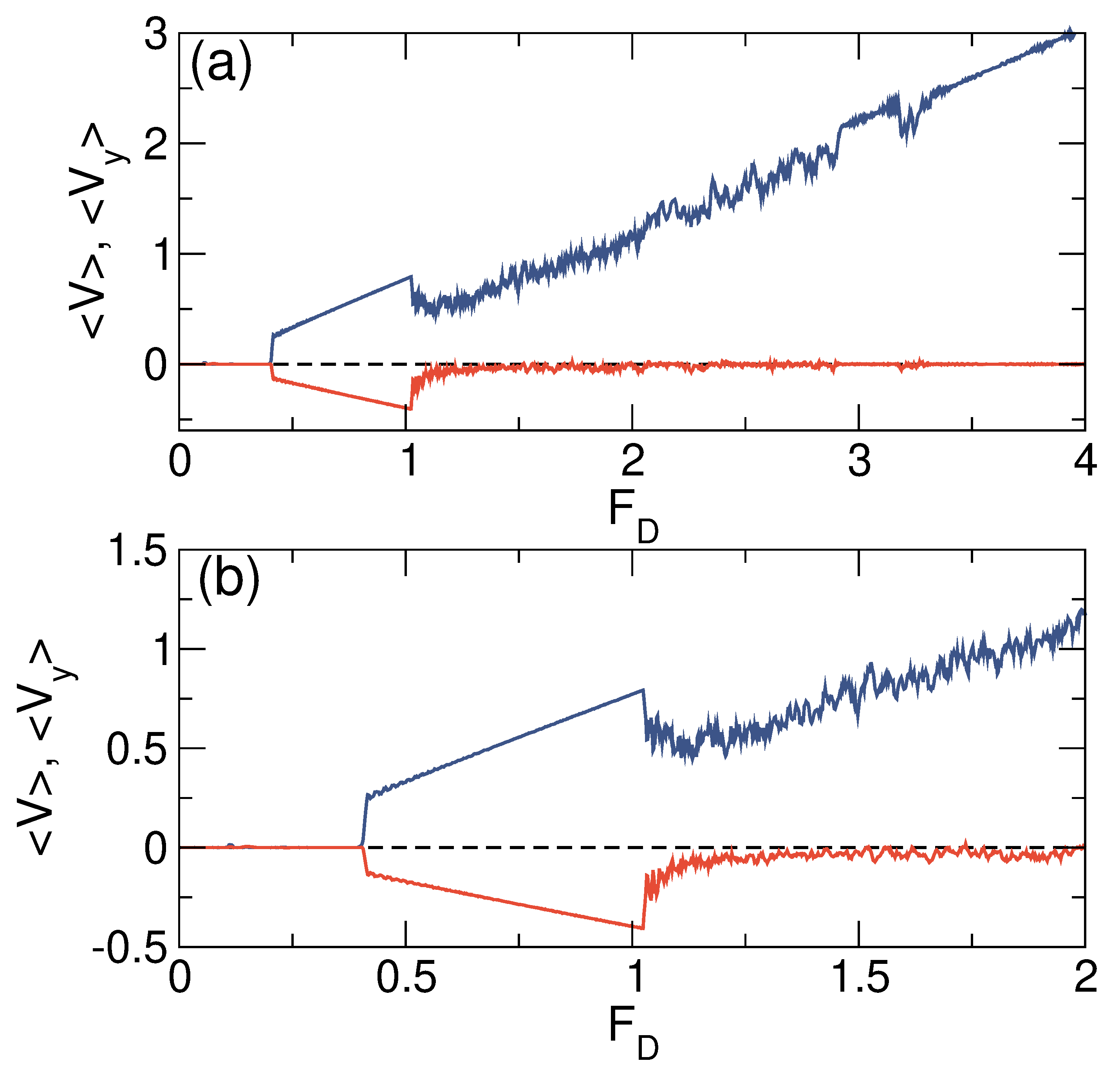}
\caption{
(a) $\langle V\rangle$ (blue) and $\langle V_{y}\rangle$ (red)
  vs $F_{D}$ for the system in Fig.~\ref{fig:12} in the stripe state
at $B=2.2$ and $\rho=0.94$, showing
a pinned phase, a Hall regime, a plastic flow regime,
and a high drive viscous flow regime.
(b) A blowup of the Hall regime from panel (a).
}
\label{fig:13}
\end{figure}

In Fig.~\ref{fig:13}(a), we plot
$\langle V\rangle$ and $\langle V_{y}\rangle$
versus $F_{D}$ for the
stripe system from Fig.~\ref{fig:12}.
The pinned regime is followed by what we term
the Hall regime, where both the $x$ and $y$ velocities increase
in magnitude with
increasing $F_D$. A blowup of the Hall regime appears
in Fig.~\ref{fig:13}(b).
At high enough drives, the probe particle begins to break through the
stripe and generate plastic rearrangements, causing a drop in
magnitude of both the
$x$ and $y$ velocity components due to the enhanced dissipation, as
illustrated in Fig.~\ref{fig:12}(b) for $F_D=1.05$.
For drives just above this breakthrough transition,
the probe particle still has a nonzero net $y$ velocity, but
$\langle V_y\rangle$ gradually approaches zero at higher drives when
the probe particle is moving fast enough to enter the viscous flow regime.
The guided Hall regime motion is absent in the
void lattice phase because the void lattice does not break symmetry
across the $y$-direction.
The particular orientation of the ordered stripe
state depends on the initial conditions and how the system is prepared.
Our results show that, in general, oriented stripes
will produce a finite Hall angle since the drive on the probe particle is
not likely to be perfectly aligned with the stripe orientation direction.
The transition to the viscous flow state at high drives appears in
Fig.~\ref{fig:13}(a) as a sudden change to a state with greatly reduced
fluctuations in $\langle V\rangle$. There is a window of viscous flow
beginning at $F_D=2.9$ that is interrupted by a temporary return to plastic
flow at $F_D=3.17$. The system then fully enters the viscous flow
state at $F_D=3.35$.

\begin{figure}
\includegraphics[width=\columnwidth]{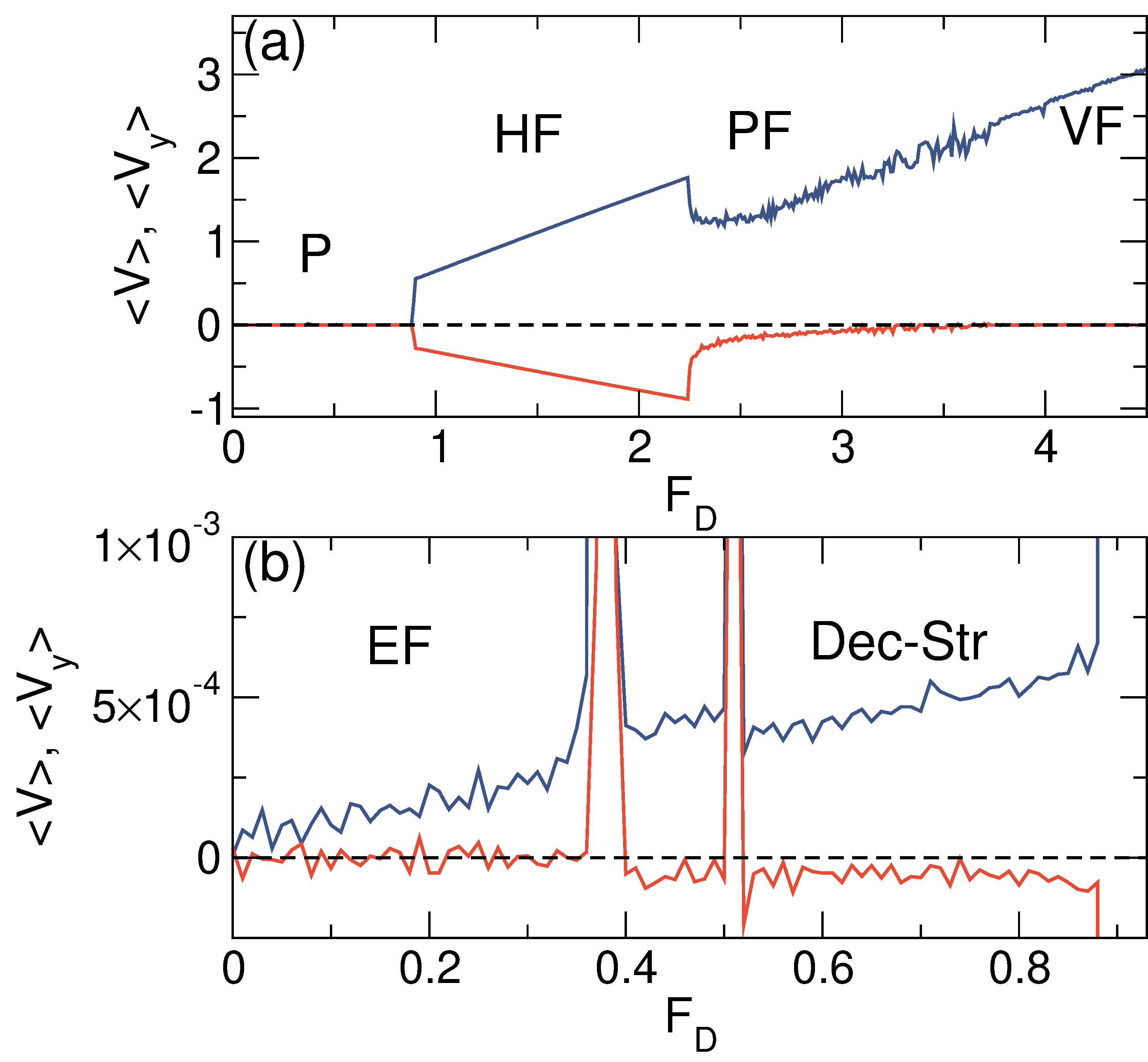}
\caption{$\langle V\rangle$ (blue) and $\langle V_y\rangle$ (red) vs $F_D$
for the oriented stripe state at $B = 2.3$ and $\rho = 1.66$.
(a) The labeled regions are a pinned (P) phase where
the probe particle remains trapped in the stripe, as
illustrated in Fig.~\ref{fig:15}(a),
the Hall flow (HF) regime shown in
Fig.~\ref{fig:15}(b),
a plastic flow (PF) phase imaged in
Fig.~\ref{fig:15}(c), and a viscous flow (VF) regime
shown in Fig.~\ref{fig:15}(d).
(b) A blowup of the pinned phase
from panel (a),
consisting of an elastic flow (EF) state
where all of the particles move very slowly together,
and a decoupled stripe (Dec-Str) state in which the stripe containing the
probe particle moves past the other stripes,
as illustrated in Fig.~\ref{fig:15}(a).
}
\label{fig:14}
\end{figure}

\begin{figure}
\includegraphics[width=\columnwidth]{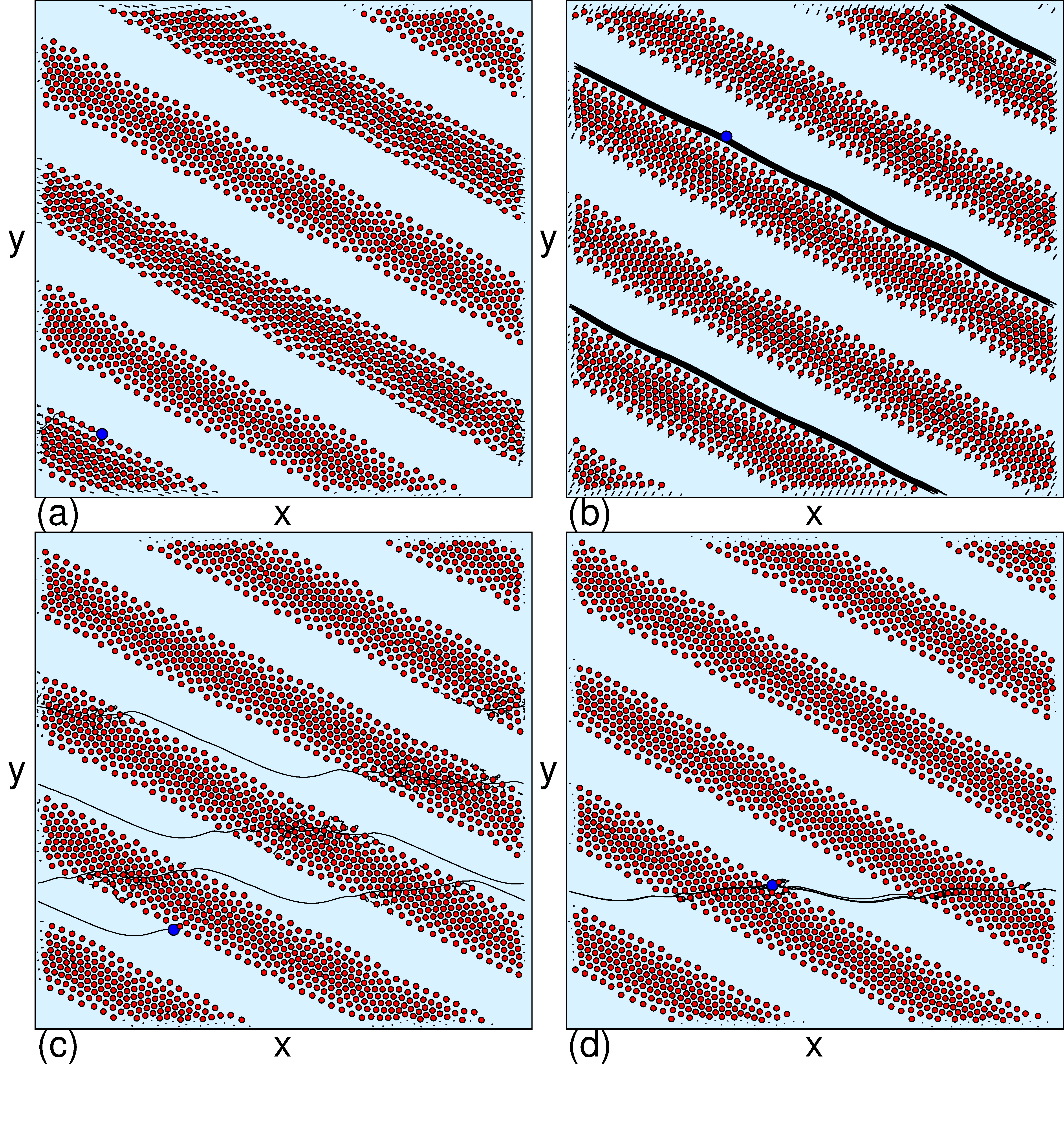}
\caption{Locations (dots) and trajectories (lines) of the probe particle
(blue) and surrounding particles (red) in the ordered stripe
state from Fig.~\ref{fig:14} with $B=2.3$ and $\rho=1.66$.
(a) At $F_{D} = 0.7$, the probe particle is trapped
inside a stripe and drags that stripe at an angle while the other
stripes do not move in the stripe decoupled phase.
(b) $F_{D} = 1.5$ in the Hall flow regime, where the probe particle
moves in the positive $x$ and negative $y$ direction,
and the stripes are dragged in the positive $y$ and positive $x$ direction.
(c) $F_{D} = 2.3$ in the plastic flow regime.
(d) $F_{D} = 4.2$ in the viscous flow regime,
where the probe particle does not create plastic deformations
as it passes through the stripes. 
}
\label{fig:15}
\end{figure}

The Hall flow regime remains robust
for other densities and $B$ values.
For example, in Fig.~\ref{fig:14}(a) we plot
$\langle V\rangle$ and $\langle V_{y}\rangle$
for a high density oriented stripe system with $B = 2.3$
and $\rho=1.66$.
The stripes are quite wide at this density, as shown in
Fig.~\ref{fig:15}.
The probe particle remains trapped in the stripe up
to $F_{D} = 0.9$, indicating that it is difficult for the probe
particle to escape from the wider stripe.
The pinned state
consists of two phases. In the first, which we call an elastic flow (EF)
regime, the entire system is dragged in the
direction of probe motion,
as shown in Fig.~\ref{fig:14}(b) where we highlight $\langle V\rangle$ and
$\langle V_y\rangle$ versus $F_D$ at drives
below the Hall regime.
Only the $x$ velocity of the probe particle is nonzero
for $F_{D} < 0.4$.
The jumps in $\langle V\rangle$ and $\langle V_y\rangle$ that
appear in Fig.~\ref{fig:14}(b) are the result of brief
large plastic rearrangements in which the probe particle shifts its position
inside the stripe but remains trapped inside the stripe.
For $F_{D} \geq 0.4$, the stripe that contains the probe particle
breaks away from the other stripes and continues to move while the other
stripes remain stationary, creating a decoupled stripe (Dec-Str)
regime. The stripe containing the probe particle moves in the positive
$x$ and negative $y$ direction, which is visible as the development of 
a finite $\langle V_y\rangle$ that grows in magnitude
in Fig.~\ref{fig:14}(b) over the range $0.4 < F_{D} < 0.9$.
The Hall angle also becomes finite at this time.
In Fig.~\ref{fig:15}(a), we
illustrate the decoupled stripe phase at $F_{D} = 0.7$
where the trajectories show that the motion occurs only in the stripe
that contains the probe particle.
In general, decoupled stripe states occur for large $B$
and large stripe widths,
where there is a more extensive range of $F_{D}$
over which the probe particle can remain trapped inside the stripe.
For $0.9 < F_{D} < 2.25$, the probe particle jumps out of the stripe and
moves along the stripe edge, as shown in
Fig.~\ref{fig:15}(b) at $F_{D} = 1.5$.
In this Hall flow regime, the probe particle moves 
in the positive $x$ and negative $y$ direction
but also drags the entire stripe in the positive $y$ and
positive $x$ direction.
For $F_{D} > 2.25$, the probe particle escapes from the stripe edge and becomes
able to pass through the stripe,
and the system enters the plastic flow regime,
illustrated at $F_D=2.3$ in Fig.~\ref{fig:15}(c).
When the probe particle breaks through the stripes,
it creates plastic rearrangements within the stripe
that reduce the velocity of the probe particle.
The system remains in the plastic flow phase
over the range $2.25 < F_{D} <  3.72$,
and when $F_{D} \geq 3.72$, a viscous flow regime
appears in which the
probe particle moves
so rapidly through the stripe that no large scale
plastic rearrangements
occur, as shown in Fig.~\ref{fig:15}(d) at $F_{D} = 4.2$.

\begin{figure}
\includegraphics[width=\columnwidth]{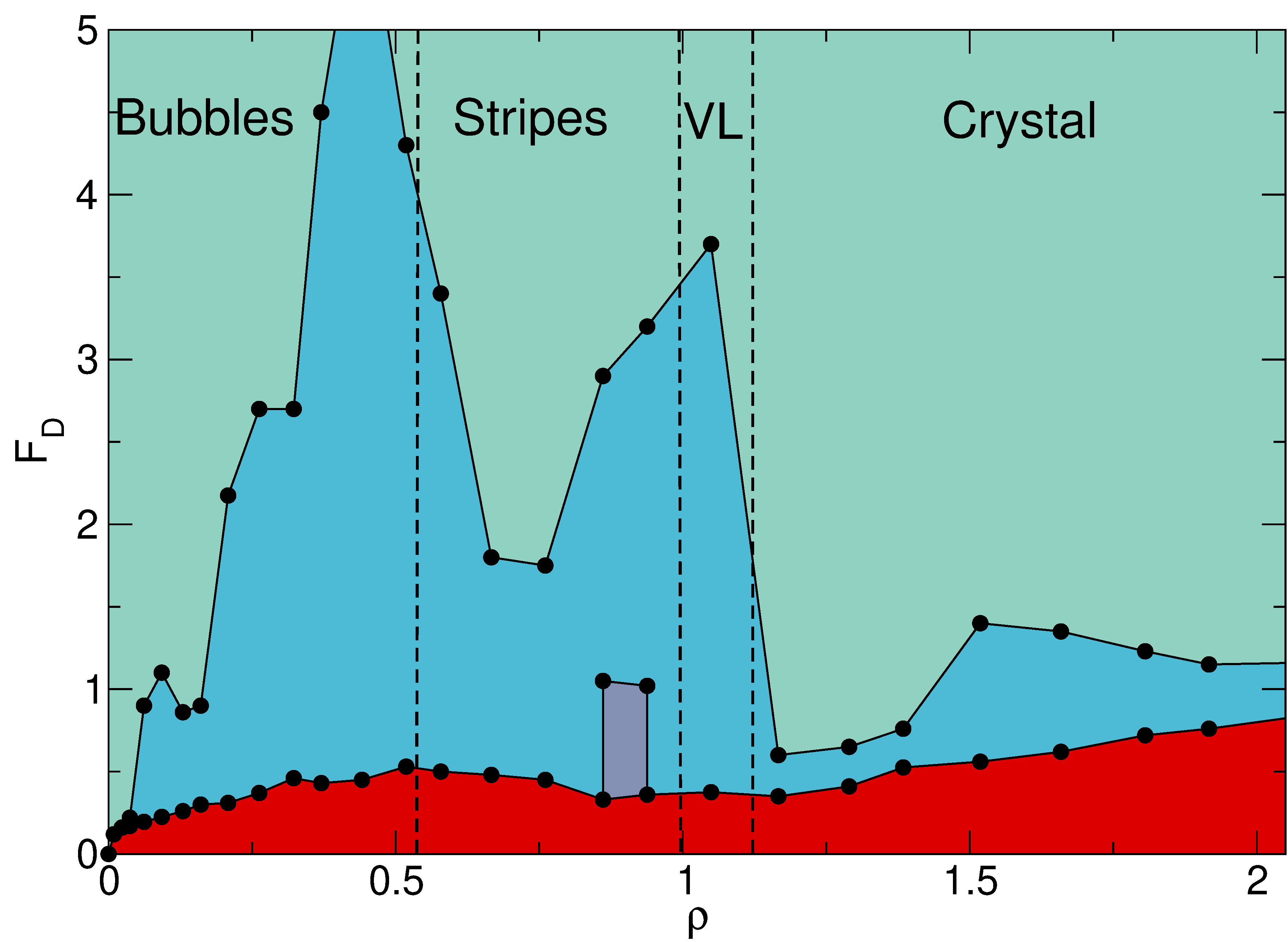}
\caption{Dynamic phase diagram as a function
of $F_{D}$  vs $\rho$  at a fixed $B = 2.2$
highlighting the pinned state
(red),
plastic flow
(blue),
viscous flow
(green),
and Hall flow
(purple).
The vertical lines
indicate the separations between the bubble, stripe, void lattice (VL),
and uniform crystal states.
}
\label{fig:16}
\end{figure}

From the velocity-force curves and the different dynamical phases,
we can construct a dynamic phase diagram highlighting
the pinned state, plastic flow, viscous flow, and Hall flow
states, as shown in
Fig.~\ref{fig:16} as a function of $F_D$ versus $\rho$ for a system with
$B = 2.2$. The vertical dashed lines
indicate the separations between the bubble, stripe, void lattice, and
uniform crystal states.
The depinning threshold
is non-monotonic as a function of $B$, as
previously shown in Fig.~\ref{fig:9}.
At low $\rho$, the transition from plastic
flow to viscous flow occurs at a very low drive,
while a local peak in the PF-VF transition appears
slightly below the bubble to stripe boundary.
As $\rho$ is increased, the bubbles grow in size and can more effectively
trap the probe particles, so a larger drive must be applied in order
to reach
the viscous flow regime.
In the stripe phase, the PF-VF transition drops to lower values because
the probe particle is traveling through stripes
that have a reduced width compared to the size of the bubbles at the upper
edge of the bubble state.
As $\rho$ increases, the width of the stripes increases,
and the extent of the plastic flow phase increases until the PF-VF
transition reaches another local maximum just above the transition to the
void lattice state.
In the uniform crystal phase, the amount of plastic flow is
greatly reduced and the PF-VF transition drops to a low value
of $F_D$. The probe particle transitions to the
viscous flow regime when it begins to
move between the rows of particles in the background crystal.
The guided Hall flow regime appears in a window where wide stripes
are present, and occurs when the stripes form an aligned state instead
of a disordered labyrinth state.

\begin{figure}
\includegraphics[width=\columnwidth]{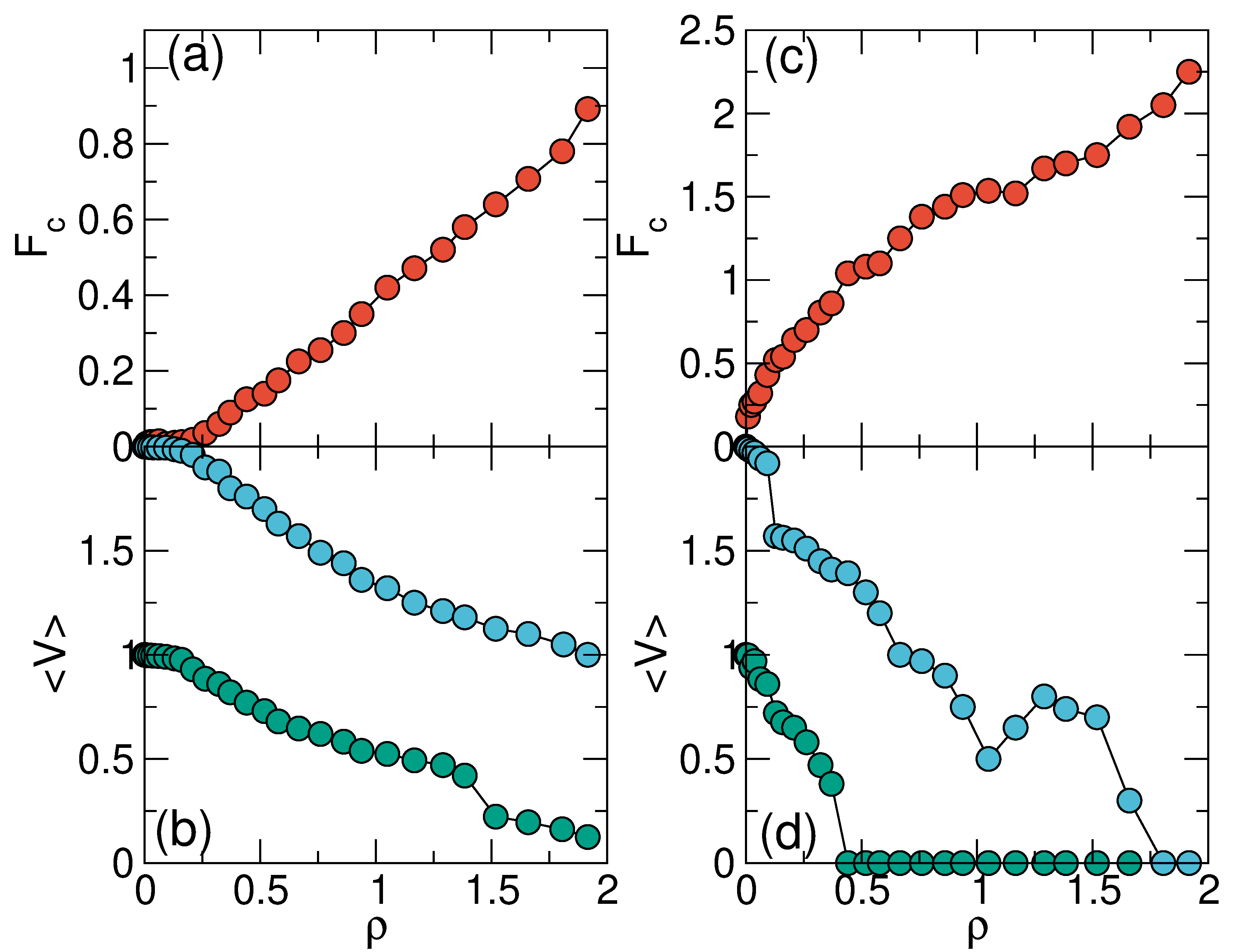}
\caption{(a) $F_{c}$ vs $\rho$ for a system with $B = 1.6$.
The system is in the uniform crystal
state for $\rho > 0.15$, and $F_c$ increases monotonically with
increasing $\rho$ in this regime.
(b) The corresponding $\langle V\rangle$ vs $\rho$
at fixed $F_{D} = 1.0$ (green) and $2.0$ (blue) shows a monotonic
decrease with $\rho$.
(c) $F_{c}$ vs $\rho$ for a system with $B = 2.4$,
where the system is in a bubble phase for all values of $\rho$.
(d) The corresponding $\langle V\rangle$ vs $\rho$ for
fixed $F_{D} = 1.0$ (green) and $2.0$ (blue).
When $F_{D} = 2.0$, the velocity has a non-monotonic dependence on $\rho$.
}
\label{fig:17}
\end{figure}

We next consider parameters for which the system exhibits
only one dominant phase.
In Fig.~\ref{fig:17}(a), we
plot $F_{c}$ versus $\rho$ for a system with $B = 1.6$,
which is in the uniform crystal phase for $\rho >  0.15$.
Throughout the uniform crystal state, $F_c$ increases monotonically
with increasing $\rho$.
Figure~\ref{fig:17}(b) shows the corresponding $\langle V\rangle$
versus $\rho$ curves
for fixed drives of $F_{D} = 1.0$ and $2.0$.
Each curve 
is flat in the low density bubble and stripe
phases,
and gradually decreases
with increasing $\rho$ in the uniform crystal phase.
At $F_{D} = 1.0$, when  $\rho > 1.4$
the probe particle is able to generate
plastic deformations in the surrounding particles, producing a
drop in $\langle V\rangle$.
In contrast, at $F_D=2.0$
the drive is high enough that the probe particle always remains in the
elastic flow regime, so there is no velocity drop at this drive.
In Fig.~\ref{fig:17}(c) we plot $F_{c}$ versus $\rho$ for the same system at
$B = 2.4$, where a bubble phase is present over the entire
range of $\rho$ values shown.
Although $F_c$ increases nearly monotonically with increasing $\rho$,
there is a plateau near $\rho = 1.0$ corresponding to
the point at which the bubbles begin to merge into elongated objects.
Figure~\ref{fig:17}(d) shows the corresponding
$\langle V\rangle$ at $F_{D} = 1.0$ and $F_{D} = 2.0$.
At both drives, when $\rho<0.1$,
the probe particle rapidly moves from bubble to
bubble in a viscous flow regime, so there are no plastic distortions.
When $F_D=1.0$, the probe particle remains trapped inside a bubble
for $\rho > 0.4$, 
while at $F_D=2.0$, the probe particle is trapped
for $\rho > 1.75$.

\section{Discussion}

In this work, the competing interaction potential we considered involved
Coulomb repulsion combined with exponential attraction;
however, many other types of competing interactions
are possible, such as a Yukawa repulsion and/or a power law attraction.
The formation of crystal, stripe, and bubble states
remains robust for a wide range of interaction potentials,
including purely repulsive potentials that have two distinct
length scales \cite{Seul95}.
Therefore, we expect many of the same active rheology effects observed here
to be relevant
for particles with other types of competing interactions, and an exploration
of alternative interactions would be an interesting direction for
future work.
We considered a dc driven probe particle,
but it would also be interesting to
study ac driven particles
to look for resonant frequency effects, or to combine dc and ac
driving to see whether ac shaking increases or decreases the effective
drag on the probe particle.
We studied the situation of constant driving force, but it is also
possible to consider a probe particle that moves with constant velocity.
In that case, the presence of different states would produce different
force fluctuations of the probe particle.
Viscous flow states would give only small force fluctuations,
while plastic flows would be associated with strong force fluctuations.
We also did not consider thermal effects.
These could be interesting since many of the phases
we observe might exhibit one creep rate
during inter-bubble or inter-stripe motion,
but have a different creep rate for hopping from stripe to stripe
or bubble to bubble.
This system is also likely to have multiple-step melting transitions,
so a probe particle subjected to constant driving would show upward or
downward jumps in 
the viscosity as the temperature is varied \cite{Reichhardt04aa}.

\section{Summary} 
We have investigated active rheology for a probe particle moving through
a pattern forming system of particles with competing long-range
repulsion and short-range attraction.
The probe particle is driven at constant force,
and we measure the threshold force
needed for motion as well as the velocity response
as the attractive term or the particle density is varied.
For a fixed density, the system forms uniform crystal, stripe,
and bubble lattice states as a function of increasing attraction.
The threshold force is a nonmonotonic function of the attractive
term and passes through
a minimum at the transition from uniform crystal to stripe.
The threshold then increases with increasing attraction
in the bubble phase.
In the uniform crystal state, the probe particle can
travel between the rows of particles,
while in the stripe and bubble states,
the probe particle can be pinned within a stripe or bubble, and upon
depinning is able to jump
from stripe to stripe or bubble to bubble.
For a fixed drive above the depinning threshold,
as the attraction is increased, the velocity has a local maximum 
at the transition from bubbles to stripes where the attractive and
repulsive forces are the most balanced and the trapping effect on the
probe particle reaches its lowest value.
We observe multiple
dynamical regimes above depinning that involve different amounts of
plastic deformations induced in the surrounding particles by the moving
probe particle.
For sufficiently large driving forces,
the probe particle moves so rapidly through the background
that the surrounding particles do not have enough time to respond via
plastic deformations, and the probe particle enters a viscous flow state
where plastic deformation is absent and the effective drag is reduced.
When the attractive term is held fixed and the particle density is increased,
we observe
transitions among bubble, stripe, void lattice, and uniform crystal
states.
The threshold for probe particle motion changes nonomonotonically
with particle density as the system passes through the different patterned
states.
For fixed driving above threshold, the effective drag is nonmonotonic, and
shows
jumps and dips across the different structural transitions.
For some parameter values, we find disordered or labyrinthine stripe
patterns, while for other parameter values, the stripes collectively
orient into an ordered state.
A probe particle moving through oriented stripes can flow easily along
the edge of the stripe, resulting in a regime of finite Hall angle motion.
At higher drives, the probe particle breaks through the stripe,
leading to a drop in the velocity or an enhancement of the drag.
We map out a dynamical phase diagram as a function of drive and density for
fixed attraction.
Our results should be relevant to a variety of
systems with competing long-range repulsion and short-range attraction
in both soft and hard condensed matter.

\begin{acknowledgements}
We gratefully acknowledge the support of the U.S. Department of
Energy through the LANL/LDRD program for this work.
This work was supported by the US Department of Energy through
the Los Alamos National Laboratory.  Los Alamos National Laboratory is
operated by Triad National Security, LLC, for the National Nuclear Security
Administration of the U. S. Department of Energy (Contract No. 892333218NCA000001).
\end{acknowledgements}

\bibliography{mybib}

\begin{thebibliography}{65}%
\makeatletter
\providecommand \@ifxundefined [1]{%
 \@ifx{#1\undefined}
}%
\providecommand \@ifnum [1]{%
 \ifnum #1\expandafter \@firstoftwo
 \else \expandafter \@secondoftwo
 \fi
}%
\providecommand \@ifx [1]{%
 \ifx #1\expandafter \@firstoftwo
 \else \expandafter \@secondoftwo
 \fi
}%
\providecommand \natexlab [1]{#1}%
\providecommand \enquote  [1]{``#1''}%
\providecommand \bibnamefont  [1]{#1}%
\providecommand \bibfnamefont [1]{#1}%
\providecommand \citenamefont [1]{#1}%
\providecommand \href@noop [0]{\@secondoftwo}%
\providecommand \href [0]{\begingroup \@sanitize@url \@href}%
\providecommand \@href[1]{\@@startlink{#1}\@@href}%
\providecommand \@@href[1]{\endgroup#1\@@endlink}%
\providecommand \@sanitize@url [0]{\catcode `\\12\catcode `\$12\catcode
  `\&12\catcode `\#12\catcode `\^12\catcode `\_12\catcode `\%12\relax}%
\providecommand \@@startlink[1]{}%
\providecommand \@@endlink[0]{}%
\providecommand \url  [0]{\begingroup\@sanitize@url \@url }%
\providecommand \@url [1]{\endgroup\@href {#1}{\urlprefix }}%
\providecommand \urlprefix  [0]{URL }%
\providecommand \Eprint [0]{\href }%
\providecommand \doibase [0]{https://doi.org/}%
\providecommand \selectlanguage [0]{\@gobble}%
\providecommand \bibinfo  [0]{\@secondoftwo}%
\providecommand \bibfield  [0]{\@secondoftwo}%
\providecommand \translation [1]{[#1]}%
\providecommand \BibitemOpen [0]{}%
\providecommand \bibitemStop [0]{}%
\providecommand \bibitemNoStop [0]{.\EOS\space}%
\providecommand \EOS [0]{\spacefactor3000\relax}%
\providecommand \BibitemShut  [1]{\csname bibitem#1\endcsname}%
\let\auto@bib@innerbib\@empty
\bibitem [{\citenamefont {Hastings}\ \emph {et~al.}(2003)\citenamefont
  {Hastings}, \citenamefont {Olson~Reichhardt},\ and\ \citenamefont
  {Reichhardt}}]{Hastings03}%
  \BibitemOpen
  \bibfield  {author} {\bibinfo {author} {\bibfnamefont {M.~B.}\ \bibnamefont
  {Hastings}}, \bibinfo {author} {\bibfnamefont {C.~J.}\ \bibnamefont
  {Olson~Reichhardt}},\ and\ \bibinfo {author} {\bibfnamefont {C.}~\bibnamefont
  {Reichhardt}},\ }\bibfield  {title} {\bibinfo {title} {Depinning by fracture
  in a glassy background},\ }\href
  {https://doi.org/10.1103/PhysRevLett.90.098302} {\bibfield  {journal}
  {\bibinfo  {journal} {Phys. Rev. Lett.}\ }\textbf {\bibinfo {volume} {90}},\
  \bibinfo {pages} {098302} (\bibinfo {year} {2003})}\BibitemShut {NoStop}%
\bibitem [{\citenamefont {Habdas}\ \emph {et~al.}(2004)\citenamefont {Habdas},
  \citenamefont {Schaar}, \citenamefont {Levitt},\ and\ \citenamefont
  {Weeks}}]{Habdas04}%
  \BibitemOpen
  \bibfield  {author} {\bibinfo {author} {\bibfnamefont {P.}~\bibnamefont
  {Habdas}}, \bibinfo {author} {\bibfnamefont {D.}~\bibnamefont {Schaar}},
  \bibinfo {author} {\bibfnamefont {A.~C.}\ \bibnamefont {Levitt}},\ and\
  \bibinfo {author} {\bibfnamefont {E.~R.}\ \bibnamefont {Weeks}},\ }\bibfield
  {title} {\bibinfo {title} {Forced motion of a probe particle near the
  colloidal glass transition},\ }\href
  {https://doi.org/10.1209/epl/i2004-10075-y} {\bibfield  {journal} {\bibinfo
  {journal} {Europhys. Lett.}\ }\textbf {\bibinfo {volume} {67}},\ \bibinfo
  {pages} {477} (\bibinfo {year} {2004})}\BibitemShut {NoStop}%
\bibitem [{\citenamefont {Reichhardt}\ and\ \citenamefont
  {Reichhardt}(2004)}]{Reichhardt04aa}%
  \BibitemOpen
  \bibfield  {author} {\bibinfo {author} {\bibfnamefont {C.}~\bibnamefont
  {Reichhardt}}\ and\ \bibinfo {author} {\bibfnamefont {C.~J.~O.}\ \bibnamefont
  {Reichhardt}},\ }\bibfield  {title} {\bibinfo {title} {Local melting and drag
  for a particle driven through a colloidal crystal},\ }\href
  {https://doi.org/10.1103/PhysRevLett.92.108301} {\bibfield  {journal}
  {\bibinfo  {journal} {Phys. Rev. Lett.}\ }\textbf {\bibinfo {volume} {92}},\
  \bibinfo {pages} {108301} (\bibinfo {year} {2004})}\BibitemShut {NoStop}%
\bibitem [{\citenamefont {Squires}\ and\ \citenamefont
  {Brady}(2005)}]{Squires05}%
  \BibitemOpen
  \bibfield  {author} {\bibinfo {author} {\bibfnamefont {T.~M.}\ \bibnamefont
  {Squires}}\ and\ \bibinfo {author} {\bibfnamefont {J.~F.}\ \bibnamefont
  {Brady}},\ }\bibfield  {title} {\bibinfo {title} {A simple paradigm for
  active and nonlinear microrheology},\ }\href
  {https://doi.org/10.1063/1.1960607} {\bibfield  {journal} {\bibinfo
  {journal} {Phys. Fluids}\ }\textbf {\bibinfo {volume} {17}},\ \bibinfo
  {pages} {073101} (\bibinfo {year} {2005})}\BibitemShut {NoStop}%
\bibitem [{\citenamefont {Gazuz}\ \emph {et~al.}(2009)\citenamefont {Gazuz},
  \citenamefont {Puertas}, \citenamefont {Voigtmann},\ and\ \citenamefont
  {Fuchs}}]{Gazuz09}%
  \BibitemOpen
  \bibfield  {author} {\bibinfo {author} {\bibfnamefont {I.}~\bibnamefont
  {Gazuz}}, \bibinfo {author} {\bibfnamefont {A.~M.}\ \bibnamefont {Puertas}},
  \bibinfo {author} {\bibfnamefont {T.}~\bibnamefont {Voigtmann}},\ and\
  \bibinfo {author} {\bibfnamefont {M.}~\bibnamefont {Fuchs}},\ }\bibfield
  {title} {\bibinfo {title} {Active and nonlinear microrheology in dense
  colloidal suspensions},\ }\href
  {https://doi.org/10.1103/PhysRevLett.102.248302} {\bibfield  {journal}
  {\bibinfo  {journal} {Phys. Rev. Lett.}\ }\textbf {\bibinfo {volume} {102}},\
  \bibinfo {pages} {248302} (\bibinfo {year} {2009})}\BibitemShut {NoStop}%
\bibitem [{\citenamefont {Khair}\ and\ \citenamefont
  {Squires}(2010)}]{Khair10}%
  \BibitemOpen
  \bibfield  {author} {\bibinfo {author} {\bibfnamefont {A.~S.}\ \bibnamefont
  {Khair}}\ and\ \bibinfo {author} {\bibfnamefont {T.~M.}\ \bibnamefont
  {Squires}},\ }\bibfield  {title} {\bibinfo {title} {Active microrheology: A
  proposed technique to measure normal stress coefficients of complex fluids},\
  }\href {https://doi.org/10.1103/PhysRevLett.105.156001} {\bibfield  {journal}
  {\bibinfo  {journal} {Phys. Rev. Lett.}\ }\textbf {\bibinfo {volume} {105}},\
  \bibinfo {pages} {156001} (\bibinfo {year} {2010})}\BibitemShut {NoStop}%
\bibitem [{\citenamefont {Winter}\ \emph {et~al.}(2012)\citenamefont {Winter},
  \citenamefont {Horbach}, \citenamefont {Virnau},\ and\ \citenamefont
  {Binder}}]{Winter12}%
  \BibitemOpen
  \bibfield  {author} {\bibinfo {author} {\bibfnamefont {D.}~\bibnamefont
  {Winter}}, \bibinfo {author} {\bibfnamefont {J.}~\bibnamefont {Horbach}},
  \bibinfo {author} {\bibfnamefont {P.}~\bibnamefont {Virnau}},\ and\ \bibinfo
  {author} {\bibfnamefont {K.}~\bibnamefont {Binder}},\ }\bibfield  {title}
  {\bibinfo {title} {Active nonlinear microrheology in a glass-forming {Y}ukawa
  fluid},\ }\href {https://doi.org/10.1103/PhysRevLett.108.028303} {\bibfield
  {journal} {\bibinfo  {journal} {Phys. Rev. Lett.}\ }\textbf {\bibinfo
  {volume} {108}},\ \bibinfo {pages} {028303} (\bibinfo {year}
  {2012})}\BibitemShut {NoStop}%
\bibitem [{\citenamefont {Anderson}\ \emph {et~al.}(2013)\citenamefont
  {Anderson}, \citenamefont {Schaar}, \citenamefont {Hentschel}, \citenamefont
  {Hay}, \citenamefont {Habdas},\ and\ \citenamefont {Weeks}}]{Anderson13}%
  \BibitemOpen
  \bibfield  {author} {\bibinfo {author} {\bibfnamefont {D.}~\bibnamefont
  {Anderson}}, \bibinfo {author} {\bibfnamefont {D.}~\bibnamefont {Schaar}},
  \bibinfo {author} {\bibfnamefont {H.~G.~E.}\ \bibnamefont {Hentschel}},
  \bibinfo {author} {\bibfnamefont {J.}~\bibnamefont {Hay}}, \bibinfo {author}
  {\bibfnamefont {P.}~\bibnamefont {Habdas}},\ and\ \bibinfo {author}
  {\bibfnamefont {E.~R.}\ \bibnamefont {Weeks}},\ }\bibfield  {title} {\bibinfo
  {title} {Local elastic response measured near the colloidal glass
  transition},\ }\href {https://doi.org/10.1063/1.4773220} {\bibfield
  {journal} {\bibinfo  {journal} {J. Chem. Phys.}\ }\textbf {\bibinfo {volume}
  {138}},\ \bibinfo {pages} {12A520} (\bibinfo {year} {2013})}\BibitemShut
  {NoStop}%
\bibitem [{\citenamefont {Swan}\ and\ \citenamefont {Zia}(2013)}]{Swan13}%
  \BibitemOpen
  \bibfield  {author} {\bibinfo {author} {\bibfnamefont {J.~W.}\ \bibnamefont
  {Swan}}\ and\ \bibinfo {author} {\bibfnamefont {R.~N.}\ \bibnamefont {Zia}},\
  }\bibfield  {title} {\bibinfo {title} {Active microrheology: fixed-velocity
  versus fixed-force},\ }\href {https://doi.org/10.1063/1.4818810} {\bibfield
  {journal} {\bibinfo  {journal} {Phys. Fluids}\ }\textbf {\bibinfo {volume}
  {25}},\ \bibinfo {pages} {083303} (\bibinfo {year} {2013})}\BibitemShut
  {NoStop}%
\bibitem [{\citenamefont {Benichou}\ \emph {et~al.}(2013)\citenamefont
  {Benichou}, \citenamefont {Illien}, \citenamefont {Mejia-Monasterio},\ and\
  \citenamefont {Oshanin}}]{Benichou13a}%
  \BibitemOpen
  \bibfield  {author} {\bibinfo {author} {\bibfnamefont {O.}~\bibnamefont
  {Benichou}}, \bibinfo {author} {\bibfnamefont {P.}~\bibnamefont {Illien}},
  \bibinfo {author} {\bibfnamefont {C.}~\bibnamefont {Mejia-Monasterio}},\ and\
  \bibinfo {author} {\bibfnamefont {G.}~\bibnamefont {Oshanin}},\ }\bibfield
  {title} {\bibinfo {title} {A biased intruder in a dense quiescent medium:
  looking beyond the force-velocity relation},\ }\href
  {https://doi.org/10.1088/1742-5468/2013/05/P05008} {\bibfield  {journal}
  {\bibinfo  {journal} {J. Stat. Mech.}\ }\textbf {\bibinfo {volume} {2013}},\
  \bibinfo {pages} {P05008} (\bibinfo {year} {2013})}\BibitemShut {NoStop}%
\bibitem [{\citenamefont {Puertas}\ and\ \citenamefont
  {Voigtmann}(2014)}]{Puertas14}%
  \BibitemOpen
  \bibfield  {author} {\bibinfo {author} {\bibfnamefont {A.~M.}\ \bibnamefont
  {Puertas}}\ and\ \bibinfo {author} {\bibfnamefont {T.}~\bibnamefont
  {Voigtmann}},\ }\bibfield  {title} {\bibinfo {title} {Microrheology of
  colloidal systems},\ }\href {https://doi.org/10.1088/0953-8984/26/24/243101}
  {\bibfield  {journal} {\bibinfo  {journal} {J. Phys.: Condens. Matter}\
  }\textbf {\bibinfo {volume} {26}},\ \bibinfo {pages} {243101} (\bibinfo
  {year} {2014})}\BibitemShut {NoStop}%
\bibitem [{\citenamefont {Gruber}\ \emph {et~al.}(2016)\citenamefont {Gruber},
  \citenamefont {Abade}, \citenamefont {Puertas},\ and\ \citenamefont
  {Fuchs}}]{Gruber16}%
  \BibitemOpen
  \bibfield  {author} {\bibinfo {author} {\bibfnamefont {M.}~\bibnamefont
  {Gruber}}, \bibinfo {author} {\bibfnamefont {G.~C.}\ \bibnamefont {Abade}},
  \bibinfo {author} {\bibfnamefont {A.~M.}\ \bibnamefont {Puertas}},\ and\
  \bibinfo {author} {\bibfnamefont {M.}~\bibnamefont {Fuchs}},\ }\bibfield
  {title} {\bibinfo {title} {Active microrheology in a colloidal glass},\
  }\href {https://doi.org/10.1103/PhysRevE.94.042602} {\bibfield  {journal}
  {\bibinfo  {journal} {Phys. Rev. E}\ }\textbf {\bibinfo {volume} {94}},\
  \bibinfo {pages} {042602} (\bibinfo {year} {2016})}\BibitemShut {NoStop}%
\bibitem [{\citenamefont {Wulfert}\ \emph {et~al.}(2017)\citenamefont
  {Wulfert}, \citenamefont {Seifert},\ and\ \citenamefont {Speck}}]{Wulfert17}%
  \BibitemOpen
  \bibfield  {author} {\bibinfo {author} {\bibfnamefont {R.}~\bibnamefont
  {Wulfert}}, \bibinfo {author} {\bibfnamefont {U.}~\bibnamefont {Seifert}},\
  and\ \bibinfo {author} {\bibfnamefont {T.}~\bibnamefont {Speck}},\ }\bibfield
   {title} {\bibinfo {title} {Nonequilibrium depletion interactions in active
  microrheology},\ }\href {https://doi.org/10.1039/C7SM01737E} {\bibfield
  {journal} {\bibinfo  {journal} {Soft Matter}\ }\textbf {\bibinfo {volume}
  {13}},\ \bibinfo {pages} {9093} (\bibinfo {year} {2017})}\BibitemShut
  {NoStop}%
\bibitem [{\citenamefont {Zia}(2018)}]{Zia18}%
  \BibitemOpen
  \bibfield  {author} {\bibinfo {author} {\bibfnamefont {R.~N.}\ \bibnamefont
  {Zia}},\ }\bibfield  {title} {\bibinfo {title} {Active and passive
  microrheology: Theory and simulation},\ }\href
  {https://doi.org/10.1146/annurev-fluid-122316-044514} {\bibfield  {journal}
  {\bibinfo  {journal} {Ann. Rev. Fluid Mech.}\ }\textbf {\bibinfo {volume}
  {50}},\ \bibinfo {pages} {371} (\bibinfo {year} {2018})}\BibitemShut
  {NoStop}%
\bibitem [{\citenamefont {Yu}\ \emph {et~al.}(2020)\citenamefont {Yu},
  \citenamefont {Rahbari}, \citenamefont {Kawasaki}, \citenamefont {Park},\
  and\ \citenamefont {Lee}}]{Yu20}%
  \BibitemOpen
  \bibfield  {author} {\bibinfo {author} {\bibfnamefont {J.~W.}\ \bibnamefont
  {Yu}}, \bibinfo {author} {\bibfnamefont {S.~H.~E.}\ \bibnamefont {Rahbari}},
  \bibinfo {author} {\bibfnamefont {T.}~\bibnamefont {Kawasaki}}, \bibinfo
  {author} {\bibfnamefont {H.}~\bibnamefont {Park}},\ and\ \bibinfo {author}
  {\bibfnamefont {W.~B.}\ \bibnamefont {Lee}},\ }\bibfield  {title} {\bibinfo
  {title} {Active microrheology of a bulk metallic glass},\ }\bibfield
  {journal} {\bibinfo  {journal} {Sci. Adv.}\ }\textbf {\bibinfo {volume}
  {6}},\ \href {https://doi.org/10.1126/sciadv.aba8766}
  {10.1126/sciadv.aba8766} (\bibinfo {year} {2020})\BibitemShut {NoStop}%
\bibitem [{\citenamefont {\c{S}enbil}\ \emph {et~al.}(2019)\citenamefont
  {\c{S}enbil}, \citenamefont {Gruber}, \citenamefont {Zhang}, \citenamefont
  {Fuchs},\ and\ \citenamefont {Scheffold}}]{Senbil19}%
  \BibitemOpen
  \bibfield  {author} {\bibinfo {author} {\bibfnamefont {N.}~\bibnamefont
  {\c{S}enbil}}, \bibinfo {author} {\bibfnamefont {M.}~\bibnamefont {Gruber}},
  \bibinfo {author} {\bibfnamefont {C.}~\bibnamefont {Zhang}}, \bibinfo
  {author} {\bibfnamefont {M.}~\bibnamefont {Fuchs}},\ and\ \bibinfo {author}
  {\bibfnamefont {F.}~\bibnamefont {Scheffold}},\ }\bibfield  {title} {\bibinfo
  {title} {Observation of strongly heterogeneous dynamics at the depinning
  transition in a colloidal glass},\ }\href
  {https://doi.org/10.1103/PhysRevLett.122.108002} {\bibfield  {journal}
  {\bibinfo  {journal} {Phys. Rev. Lett.}\ }\textbf {\bibinfo {volume} {122}},\
  \bibinfo {pages} {108002} (\bibinfo {year} {2019})}\BibitemShut {NoStop}%
\bibitem [{\citenamefont {Gruber}\ \emph {et~al.}(2020)\citenamefont {Gruber},
  \citenamefont {Puertas},\ and\ \citenamefont {Fuchs}}]{Gruber20}%
  \BibitemOpen
  \bibfield  {author} {\bibinfo {author} {\bibfnamefont {M.}~\bibnamefont
  {Gruber}}, \bibinfo {author} {\bibfnamefont {A.~M.}\ \bibnamefont
  {Puertas}},\ and\ \bibinfo {author} {\bibfnamefont {M.}~\bibnamefont
  {Fuchs}},\ }\bibfield  {title} {\bibinfo {title} {Critical force in active
  microrheology},\ }\href {https://doi.org/10.1103/PhysRevE.101.012612}
  {\bibfield  {journal} {\bibinfo  {journal} {Phys. Rev. E}\ }\textbf {\bibinfo
  {volume} {101}},\ \bibinfo {pages} {012612} (\bibinfo {year}
  {2020})}\BibitemShut {NoStop}%
\bibitem [{\citenamefont {Hopkins}\ \emph {et~al.}(2022)\citenamefont
  {Hopkins}, \citenamefont {Chiang}, \citenamefont {Loewe}, \citenamefont
  {Marenduzzo},\ and\ \citenamefont {Marchetti}}]{Hopkins22}%
  \BibitemOpen
  \bibfield  {author} {\bibinfo {author} {\bibfnamefont {A.}~\bibnamefont
  {Hopkins}}, \bibinfo {author} {\bibfnamefont {M.}~\bibnamefont {Chiang}},
  \bibinfo {author} {\bibfnamefont {B.}~\bibnamefont {Loewe}}, \bibinfo
  {author} {\bibfnamefont {D.}~\bibnamefont {Marenduzzo}},\ and\ \bibinfo
  {author} {\bibfnamefont {M.~C.}\ \bibnamefont {Marchetti}},\ }\bibfield
  {title} {\bibinfo {title} {Local yield and compliance in active cell
  monolayers},\ }\href {https://doi.org/10.1103/PhysRevLett.129.148101}
  {\bibfield  {journal} {\bibinfo  {journal} {Phys. Rev. Lett.}\ }\textbf
  {\bibinfo {volume} {129}},\ \bibinfo {pages} {148101} (\bibinfo {year}
  {2022})}\BibitemShut {NoStop}%
\bibitem [{\citenamefont {Drocco}\ \emph {et~al.}(2005)\citenamefont {Drocco},
  \citenamefont {Hastings}, \citenamefont {Reichhardt},\ and\ \citenamefont
  {Reichhardt}}]{Drocco05}%
  \BibitemOpen
  \bibfield  {author} {\bibinfo {author} {\bibfnamefont {J.~A.}\ \bibnamefont
  {Drocco}}, \bibinfo {author} {\bibfnamefont {M.~B.}\ \bibnamefont
  {Hastings}}, \bibinfo {author} {\bibfnamefont {C.~J.~O.}\ \bibnamefont
  {Reichhardt}},\ and\ \bibinfo {author} {\bibfnamefont {C.}~\bibnamefont
  {Reichhardt}},\ }\bibfield  {title} {\bibinfo {title} {Multiscaling at point
  {$J$}: Jamming is a critical phenomenon},\ }\href
  {https://doi.org/10.1103/PhysRevLett.95.088001} {\bibfield  {journal}
  {\bibinfo  {journal} {Phys. Rev. Lett.}\ }\textbf {\bibinfo {volume} {95}},\
  \bibinfo {pages} {088001} (\bibinfo {year} {2005})}\BibitemShut {NoStop}%
\bibitem [{\citenamefont {Candelier}\ and\ \citenamefont
  {Dauchot}(2010)}]{Candelier10}%
  \BibitemOpen
  \bibfield  {author} {\bibinfo {author} {\bibfnamefont {R.}~\bibnamefont
  {Candelier}}\ and\ \bibinfo {author} {\bibfnamefont {O.}~\bibnamefont
  {Dauchot}},\ }\bibfield  {title} {\bibinfo {title} {Journey of an intruder
  through the fluidization and jamming transitions of a dense granular media},\
  }\href {https://doi.org/10.1103/PhysRevE.81.011304} {\bibfield  {journal}
  {\bibinfo  {journal} {Phys. Rev. E}\ }\textbf {\bibinfo {volume} {81}},\
  \bibinfo {pages} {011304} (\bibinfo {year} {2010})}\BibitemShut {NoStop}%
\bibitem [{\citenamefont {Kolb}\ \emph {et~al.}(2013)\citenamefont {Kolb},
  \citenamefont {Cixous}, \citenamefont {Gaudouen},\ and\ \citenamefont
  {Darnige}}]{Kolb13}%
  \BibitemOpen
  \bibfield  {author} {\bibinfo {author} {\bibfnamefont {E.}~\bibnamefont
  {Kolb}}, \bibinfo {author} {\bibfnamefont {P.}~\bibnamefont {Cixous}},
  \bibinfo {author} {\bibfnamefont {N.}~\bibnamefont {Gaudouen}},\ and\
  \bibinfo {author} {\bibfnamefont {T.}~\bibnamefont {Darnige}},\ }\bibfield
  {title} {\bibinfo {title} {Rigid intruder inside a two-dimensional dense
  granular flow: Drag force and cavity formation},\ }\href
  {https://doi.org/10.1103/PhysRevE.87.032207} {\bibfield  {journal} {\bibinfo
  {journal} {Phys. Rev. E}\ }\textbf {\bibinfo {volume} {87}},\ \bibinfo
  {pages} {032207} (\bibinfo {year} {2013})}\BibitemShut {NoStop}%
\bibitem [{\citenamefont {Reichhardt}\ and\ \citenamefont
  {Reichhardt}(2019)}]{Reichhardt19a}%
  \BibitemOpen
  \bibfield  {author} {\bibinfo {author} {\bibfnamefont {C.}~\bibnamefont
  {Reichhardt}}\ and\ \bibinfo {author} {\bibfnamefont {C.~J.~O.}\ \bibnamefont
  {Reichhardt}},\ }\bibfield  {title} {\bibinfo {title} {Active microrheology,
  {H}all effect, and jamming in chiral fluids},\ }\href
  {https://doi.org/10.1103/PhysRevE.100.012604} {\bibfield  {journal} {\bibinfo
   {journal} {Phys. Rev. E}\ }\textbf {\bibinfo {volume} {100}},\ \bibinfo
  {pages} {012604} (\bibinfo {year} {2019})}\BibitemShut {NoStop}%
\bibitem [{\citenamefont {Duclut}\ \emph {et~al.}(2024)\citenamefont {Duclut},
  \citenamefont {Bo}, \citenamefont {Lier}, \citenamefont {Armas},
  \citenamefont {Sur\'owka},\ and\ \citenamefont {J\"ulicher}}]{Duclut24}%
  \BibitemOpen
  \bibfield  {author} {\bibinfo {author} {\bibfnamefont {C.}~\bibnamefont
  {Duclut}}, \bibinfo {author} {\bibfnamefont {S.}~\bibnamefont {Bo}}, \bibinfo
  {author} {\bibfnamefont {R.}~\bibnamefont {Lier}}, \bibinfo {author}
  {\bibfnamefont {J.}~\bibnamefont {Armas}}, \bibinfo {author} {\bibfnamefont
  {P.}~\bibnamefont {Sur\'owka}},\ and\ \bibinfo {author} {\bibfnamefont
  {F.}~\bibnamefont {J\"ulicher}},\ }\bibfield  {title} {\bibinfo {title}
  {Probe particles in odd active viscoelastic fluids: How activity and
  dissipation determine linear stability},\ }\href
  {https://doi.org/10.1103/PhysRevE.109.044126} {\bibfield  {journal} {\bibinfo
   {journal} {Phys. Rev. E}\ }\textbf {\bibinfo {volume} {109}},\ \bibinfo
  {pages} {044126} (\bibinfo {year} {2024})}\BibitemShut {NoStop}%
\bibitem [{\citenamefont {Reichhardt}\ and\ \citenamefont
  {Reichhardt}(2015)}]{Reichhardt15}%
  \BibitemOpen
  \bibfield  {author} {\bibinfo {author} {\bibfnamefont {C.}~\bibnamefont
  {Reichhardt}}\ and\ \bibinfo {author} {\bibfnamefont {C.~J.~O.}\ \bibnamefont
  {Reichhardt}},\ }\bibfield  {title} {\bibinfo {title} {Active microrheology
  in active matter systems: Mobility, intermittency, and avalanches},\ }\href
  {https://doi.org/10.1103/PhysRevE.91.032313} {\bibfield  {journal} {\bibinfo
  {journal} {Phys. Rev. E}\ }\textbf {\bibinfo {volume} {91}},\ \bibinfo
  {pages} {032313} (\bibinfo {year} {2015})}\BibitemShut {NoStop}%
\bibitem [{\citenamefont {Straver}\ \emph {et~al.}(2008)\citenamefont
  {Straver}, \citenamefont {Hoffman}, \citenamefont {Auslaender}, \citenamefont
  {Rugar},\ and\ \citenamefont {Moler}}]{Straver08}%
  \BibitemOpen
  \bibfield  {author} {\bibinfo {author} {\bibfnamefont {E.~W.~J.}\
  \bibnamefont {Straver}}, \bibinfo {author} {\bibfnamefont {J.~E.}\
  \bibnamefont {Hoffman}}, \bibinfo {author} {\bibfnamefont {O.~M.}\
  \bibnamefont {Auslaender}}, \bibinfo {author} {\bibfnamefont
  {D.}~\bibnamefont {Rugar}},\ and\ \bibinfo {author} {\bibfnamefont {K.~A.}\
  \bibnamefont {Moler}},\ }\bibfield  {title} {\bibinfo {title} {Controlled
  manipulation of individual vortices in a superconductor},\ }\href
  {https://doi.org/10.1063/1.3000963} {\bibfield  {journal} {\bibinfo
  {journal} {Appl. Phys. Lett.}\ }\textbf {\bibinfo {volume} {93}},\ \bibinfo
  {pages} {172514} (\bibinfo {year} {2008})}\BibitemShut {NoStop}%
\bibitem [{\citenamefont {Auslaender}\ \emph {et~al.}(2009)\citenamefont
  {Auslaender}, \citenamefont {Luan}, \citenamefont {Straver}, \citenamefont
  {Hoffman}, \citenamefont {Koshnick}, \citenamefont {Zeldov}, \citenamefont
  {Bonn}, \citenamefont {Liang}, \citenamefont {Hardy},\ and\ \citenamefont
  {Moler}}]{Auslaender09}%
  \BibitemOpen
  \bibfield  {author} {\bibinfo {author} {\bibfnamefont {O.~M.}\ \bibnamefont
  {Auslaender}}, \bibinfo {author} {\bibfnamefont {L.}~\bibnamefont {Luan}},
  \bibinfo {author} {\bibfnamefont {E.~W.~J.}\ \bibnamefont {Straver}},
  \bibinfo {author} {\bibfnamefont {J.~E.}\ \bibnamefont {Hoffman}}, \bibinfo
  {author} {\bibfnamefont {N.~C.}\ \bibnamefont {Koshnick}}, \bibinfo {author}
  {\bibfnamefont {E.}~\bibnamefont {Zeldov}}, \bibinfo {author} {\bibfnamefont
  {D.~A.}\ \bibnamefont {Bonn}}, \bibinfo {author} {\bibfnamefont
  {R.}~\bibnamefont {Liang}}, \bibinfo {author} {\bibfnamefont {W.~N.}\
  \bibnamefont {Hardy}},\ and\ \bibinfo {author} {\bibfnamefont {K.~A.}\
  \bibnamefont {Moler}},\ }\bibfield  {title} {\bibinfo {title} {Mechanics of
  individual isolated vortices in a cuprate superconductor},\ }\href
  {https://doi.org/10.1038/NPHYS1127} {\bibfield  {journal} {\bibinfo
  {journal} {Nature Phys.}\ }\textbf {\bibinfo {volume} {5}},\ \bibinfo {pages}
  {35} (\bibinfo {year} {2009})}\BibitemShut {NoStop}%
\bibitem [{\citenamefont {Reichhardt}(2009)}]{Reichhardt09a}%
  \BibitemOpen
  \bibfield  {author} {\bibinfo {author} {\bibfnamefont {C.}~\bibnamefont
  {Reichhardt}},\ }\bibfield  {title} {\bibinfo {title} {Vortices wiggled and
  dragged},\ }\href {https://doi.org/10.1038/nphys1169} {\bibfield  {journal}
  {\bibinfo  {journal} {Nature Phys.}\ }\textbf {\bibinfo {volume} {5}},\
  \bibinfo {pages} {15} (\bibinfo {year} {2009})}\BibitemShut {NoStop}%
\bibitem [{\citenamefont {Wang}\ \emph {et~al.}(2020)\citenamefont {Wang},
  \citenamefont {Chotorlishvili}, \citenamefont {Dugaev}, \citenamefont
  {Ernst}, \citenamefont {Maznichenko}, \citenamefont {Arnold}, \citenamefont
  {Jia}, \citenamefont {Berakdar}, \citenamefont {Mertig},\ and\ \citenamefont
  {Barna{\` s}}}]{Wang20b}%
  \BibitemOpen
  \bibfield  {author} {\bibinfo {author} {\bibfnamefont {X.-G.}\ \bibnamefont
  {Wang}}, \bibinfo {author} {\bibfnamefont {L.}~\bibnamefont
  {Chotorlishvili}}, \bibinfo {author} {\bibfnamefont {V.~K.}\ \bibnamefont
  {Dugaev}}, \bibinfo {author} {\bibfnamefont {A.}~\bibnamefont {Ernst}},
  \bibinfo {author} {\bibfnamefont {I.~V.}\ \bibnamefont {Maznichenko}},
  \bibinfo {author} {\bibfnamefont {N.}~\bibnamefont {Arnold}}, \bibinfo
  {author} {\bibfnamefont {C.}~\bibnamefont {Jia}}, \bibinfo {author}
  {\bibfnamefont {J.}~\bibnamefont {Berakdar}}, \bibinfo {author}
  {\bibfnamefont {I.}~\bibnamefont {Mertig}},\ and\ \bibinfo {author}
  {\bibfnamefont {J.}~\bibnamefont {Barna{\` s}}},\ }\bibfield  {title}
  {\bibinfo {title} {The optical tweezer of skyrmions},\ }\href
  {https://doi.org/10.1038/s41524-020-00402-7} {\bibfield  {journal} {\bibinfo
  {journal} {npj Comput. Mater.}\ }\textbf {\bibinfo {volume} {6}},\ \bibinfo
  {pages} {140} (\bibinfo {year} {2020})}\BibitemShut {NoStop}%
\bibitem [{\citenamefont {Reichhardt}\ and\ \citenamefont
  {Reichhardt}(2021)}]{Reichhardt21a}%
  \BibitemOpen
  \bibfield  {author} {\bibinfo {author} {\bibfnamefont {C.}~\bibnamefont
  {Reichhardt}}\ and\ \bibinfo {author} {\bibfnamefont {C.~J.~O.}\ \bibnamefont
  {Reichhardt}},\ }\bibfield  {title} {\bibinfo {title} {Dynamics and
  nonmonotonic drag for individually driven skyrmions},\ }\href
  {https://doi.org/10.1103/PhysRevB.104.064441} {\bibfield  {journal} {\bibinfo
   {journal} {Phys. Rev. B}\ }\textbf {\bibinfo {volume} {104}},\ \bibinfo
  {pages} {064441} (\bibinfo {year} {2021})}\BibitemShut {NoStop}%
\bibitem [{\citenamefont {Seul}\ and\ \citenamefont {Andelman}(1995)}]{Seul95}%
  \BibitemOpen
  \bibfield  {author} {\bibinfo {author} {\bibfnamefont {M.}~\bibnamefont
  {Seul}}\ and\ \bibinfo {author} {\bibfnamefont {D.}~\bibnamefont
  {Andelman}},\ }\bibfield  {title} {\bibinfo {title} {Domain shapes and
  patterns - the phenomenology of modulated phases},\ }\href
  {https://doi.org/10.1126/science.267.5197.476} {\bibfield  {journal}
  {\bibinfo  {journal} {Science}\ }\textbf {\bibinfo {volume} {267}},\ \bibinfo
  {pages} {476} (\bibinfo {year} {1995})}\BibitemShut {NoStop}%
\bibitem [{\citenamefont {Stoycheva}\ and\ \citenamefont
  {Singer}(2000)}]{Stoycheva00}%
  \BibitemOpen
  \bibfield  {author} {\bibinfo {author} {\bibfnamefont {A.~D.}\ \bibnamefont
  {Stoycheva}}\ and\ \bibinfo {author} {\bibfnamefont {S.~J.}\ \bibnamefont
  {Singer}},\ }\bibfield  {title} {\bibinfo {title} {Stripe melting in a
  two-dimensional system with competing interactions},\ }\href
  {https://doi.org/10.1103/PhysRevLett.84.4657} {\bibfield  {journal} {\bibinfo
   {journal} {Phys. Rev. Lett.}\ }\textbf {\bibinfo {volume} {84}},\ \bibinfo
  {pages} {4657} (\bibinfo {year} {2000})}\BibitemShut {NoStop}%
\bibitem [{\citenamefont {Reichhardt}\ \emph
  {et~al.}(2003{\natexlab{a}})\citenamefont {Reichhardt}, \citenamefont
  {Olson}, \citenamefont {Martin},\ and\ \citenamefont
  {Bishop}}]{Reichhardt03}%
  \BibitemOpen
  \bibfield  {author} {\bibinfo {author} {\bibfnamefont {C.}~\bibnamefont
  {Reichhardt}}, \bibinfo {author} {\bibfnamefont {C.~J.}\ \bibnamefont
  {Olson}}, \bibinfo {author} {\bibfnamefont {I.}~\bibnamefont {Martin}},\ and\
  \bibinfo {author} {\bibfnamefont {A.~R.}\ \bibnamefont {Bishop}},\ }\bibfield
   {title} {\bibinfo {title} {Depinning and dynamics of systems with competing
  interactions in quenched disorder},\ }\href
  {https://doi.org/10.1209/epl/i2003-00222-0} {\bibfield  {journal} {\bibinfo
  {journal} {Europhys. Lett.}\ }\textbf {\bibinfo {volume} {61}},\ \bibinfo
  {pages} {221} (\bibinfo {year} {2003}{\natexlab{a}})}\BibitemShut {NoStop}%
\bibitem [{\citenamefont {Reichhardt}\ \emph {et~al.}(2004)\citenamefont
  {Reichhardt}, \citenamefont {Reichhardt}, \citenamefont {Martin},\ and\
  \citenamefont {Bishop}}]{Reichhardt04}%
  \BibitemOpen
  \bibfield  {author} {\bibinfo {author} {\bibfnamefont {C.~J.~O.}\
  \bibnamefont {Reichhardt}}, \bibinfo {author} {\bibfnamefont
  {C.}~\bibnamefont {Reichhardt}}, \bibinfo {author} {\bibfnamefont
  {I.}~\bibnamefont {Martin}},\ and\ \bibinfo {author} {\bibfnamefont {A.~R.}\
  \bibnamefont {Bishop}},\ }\bibfield  {title} {\bibinfo {title} {Dynamics and
  melting of stripes, crystals, and bubbles with quenched disorder},\ }\href
  {https://doi.org/10.1016/j.physd.2004.01.027} {\bibfield  {journal} {\bibinfo
   {journal} {Physica D}\ }\textbf {\bibinfo {volume} {193}},\ \bibinfo {pages}
  {303} (\bibinfo {year} {2004})}\BibitemShut {NoStop}%
\bibitem [{\citenamefont {Mossa}\ \emph {et~al.}(2004)\citenamefont {Mossa},
  \citenamefont {Sciortino}, \citenamefont {Tartaglia},\ and\ \citenamefont
  {Zaccarelli}}]{Mossa04}%
  \BibitemOpen
  \bibfield  {author} {\bibinfo {author} {\bibfnamefont {S.}~\bibnamefont
  {Mossa}}, \bibinfo {author} {\bibfnamefont {F.}~\bibnamefont {Sciortino}},
  \bibinfo {author} {\bibfnamefont {P.}~\bibnamefont {Tartaglia}},\ and\
  \bibinfo {author} {\bibfnamefont {E.}~\bibnamefont {Zaccarelli}},\ }\bibfield
   {title} {\bibinfo {title} {Ground-state clusters for short-range attractive
  and long-range repulsive potentials},\ }\href
  {https://doi.org/10.1021/la048554t} {\bibfield  {journal} {\bibinfo
  {journal} {Langmuir}\ }\textbf {\bibinfo {volume} {20}},\ \bibinfo {pages}
  {10756} (\bibinfo {year} {2004})}\BibitemShut {NoStop}%
\bibitem [{\citenamefont {Sciortino}\ \emph {et~al.}(2004)\citenamefont
  {Sciortino}, \citenamefont {Mossa}, \citenamefont {Zaccarelli},\ and\
  \citenamefont {Tartaglia}}]{Sciortino04}%
  \BibitemOpen
  \bibfield  {author} {\bibinfo {author} {\bibfnamefont {F.}~\bibnamefont
  {Sciortino}}, \bibinfo {author} {\bibfnamefont {S.}~\bibnamefont {Mossa}},
  \bibinfo {author} {\bibfnamefont {E.}~\bibnamefont {Zaccarelli}},\ and\
  \bibinfo {author} {\bibfnamefont {P.}~\bibnamefont {Tartaglia}},\ }\bibfield
  {title} {\bibinfo {title} {Equilibrium cluster phases and low-density
  arrested disordered states: The role of short-range attraction and long-range
  repulsion},\ }\href {https://doi.org/10.1103/PhysRevLett.93.055701}
  {\bibfield  {journal} {\bibinfo  {journal} {Phys. Rev. Lett.}\ }\textbf
  {\bibinfo {volume} {93}},\ \bibinfo {pages} {055701} (\bibinfo {year}
  {2004})}\BibitemShut {NoStop}%
\bibitem [{\citenamefont {Nelissen}\ \emph {et~al.}(2005)\citenamefont
  {Nelissen}, \citenamefont {Partoens},\ and\ \citenamefont
  {Peeters}}]{Nelissen05}%
  \BibitemOpen
  \bibfield  {author} {\bibinfo {author} {\bibfnamefont {K.}~\bibnamefont
  {Nelissen}}, \bibinfo {author} {\bibfnamefont {B.}~\bibnamefont {Partoens}},\
  and\ \bibinfo {author} {\bibfnamefont {F.~M.}\ \bibnamefont {Peeters}},\
  }\bibfield  {title} {\bibinfo {title} {Bubble, stripe, and ring phases in a
  two-dimensional cluster with competing interactions},\ }\href
  {https://doi.org/10.1103/PhysRevE.71.066204} {\bibfield  {journal} {\bibinfo
  {journal} {Phys. Rev. E}\ }\textbf {\bibinfo {volume} {71}},\ \bibinfo
  {pages} {066204} (\bibinfo {year} {2005})}\BibitemShut {NoStop}%
\bibitem [{\citenamefont {Liu}\ \emph {et~al.}(2008)\citenamefont {Liu},
  \citenamefont {Chew},\ and\ \citenamefont {Yu}}]{Liu08}%
  \BibitemOpen
  \bibfield  {author} {\bibinfo {author} {\bibfnamefont {Y.~H.}\ \bibnamefont
  {Liu}}, \bibinfo {author} {\bibfnamefont {L.~Y.}\ \bibnamefont {Chew}},\ and\
  \bibinfo {author} {\bibfnamefont {M.~Y.}\ \bibnamefont {Yu}},\ }\bibfield
  {title} {\bibinfo {title} {Self-assembly of complex structures in a
  two-dimensional system with competing interaction forces},\ }\href
  {https://doi.org/10.1103/PhysRevE.78.066405} {\bibfield  {journal} {\bibinfo
  {journal} {Phys. Rev. E}\ }\textbf {\bibinfo {volume} {78}},\ \bibinfo
  {pages} {066405} (\bibinfo {year} {2008})}\BibitemShut {NoStop}%
\bibitem [{\citenamefont {Olson~Reichhardt}\ \emph {et~al.}(2010)\citenamefont
  {Olson~Reichhardt}, \citenamefont {Reichhardt},\ and\ \citenamefont
  {Bishop}}]{Reichhardt10}%
  \BibitemOpen
  \bibfield  {author} {\bibinfo {author} {\bibfnamefont {C.~J.}\ \bibnamefont
  {Olson~Reichhardt}}, \bibinfo {author} {\bibfnamefont {C.}~\bibnamefont
  {Reichhardt}},\ and\ \bibinfo {author} {\bibfnamefont {A.~R.}\ \bibnamefont
  {Bishop}},\ }\bibfield  {title} {\bibinfo {title} {Structural transitions,
  melting, and intermediate phases for stripe- and clump-forming systems},\
  }\href {https://doi.org/10.1103/PhysRevE.82.041502} {\bibfield  {journal}
  {\bibinfo  {journal} {Phys. Rev. E}\ }\textbf {\bibinfo {volume} {82}},\
  \bibinfo {pages} {041502} (\bibinfo {year} {2010})}\BibitemShut {NoStop}%
\bibitem [{\citenamefont {McDermott}\ \emph {et~al.}(2014)\citenamefont
  {McDermott}, \citenamefont {Reichhardt},\ and\ \citenamefont
  {Reichhardt}}]{McDermott14}%
  \BibitemOpen
  \bibfield  {author} {\bibinfo {author} {\bibfnamefont {D.}~\bibnamefont
  {McDermott}}, \bibinfo {author} {\bibfnamefont {C.~J.~O.}\ \bibnamefont
  {Reichhardt}},\ and\ \bibinfo {author} {\bibfnamefont {C.}~\bibnamefont
  {Reichhardt}},\ }\bibfield  {title} {\bibinfo {title} {Stripe systems with
  competing interactions on quasi-one dimensional periodic substrates},\ }\href
  {https://doi.org/10.1039/c4sm01341g} {\bibfield  {journal} {\bibinfo
  {journal} {Soft Matter}\ }\textbf {\bibinfo {volume} {10}},\ \bibinfo {pages}
  {6332} (\bibinfo {year} {2014})}\BibitemShut {NoStop}%
\bibitem [{\citenamefont {Liu}\ and\ \citenamefont {Xi}(2019)}]{Liu19}%
  \BibitemOpen
  \bibfield  {author} {\bibinfo {author} {\bibfnamefont {Y.}~\bibnamefont
  {Liu}}\ and\ \bibinfo {author} {\bibfnamefont {Y.}~\bibnamefont {Xi}},\
  }\bibfield  {title} {\bibinfo {title} {Colloidal systems with a short-range
  attraction and long-range repulsion: phase diagrams, structures, and
  dynamics},\ }\href {https://doi.org/10.1016/j.cocis.2019.01.016} {\bibfield
  {journal} {\bibinfo  {journal} {Curr. Opin. Colloid Interf. Sci.}\ }\textbf
  {\bibinfo {volume} {19}},\ \bibinfo {pages} {123} (\bibinfo {year}
  {2019})}\BibitemShut {NoStop}%
\bibitem [{\citenamefont {Al~Harraq}\ \emph {et~al.}(2022)\citenamefont
  {Al~Harraq}, \citenamefont {Hymel}, \citenamefont {Lin}, \citenamefont
  {Truskett},\ and\ \citenamefont {Bharti}}]{AlHarraq22}%
  \BibitemOpen
  \bibfield  {author} {\bibinfo {author} {\bibfnamefont {A.}~\bibnamefont
  {Al~Harraq}}, \bibinfo {author} {\bibfnamefont {A.~A.}\ \bibnamefont
  {Hymel}}, \bibinfo {author} {\bibfnamefont {E.}~\bibnamefont {Lin}}, \bibinfo
  {author} {\bibfnamefont {T.~M.}\ \bibnamefont {Truskett}},\ and\ \bibinfo
  {author} {\bibfnamefont {B.}~\bibnamefont {Bharti}},\ }\bibfield  {title}
  {\bibinfo {title} {Dual nature of magnetic nanoparticle dispersions enables
  control over short-range attraction and long-range repulsion interactions},\
  }\href {https://doi.org/10.1038/s42004-022-00687-3} {\bibfield  {journal}
  {\bibinfo  {journal} {Commun. Chem.}\ }\textbf {\bibinfo {volume} {5}},\
  \bibinfo {pages} {72} (\bibinfo {year} {2022})}\BibitemShut {NoStop}%
\bibitem [{\citenamefont {Hooshanginejad}\ \emph {et~al.}(2024)\citenamefont
  {Hooshanginejad}, \citenamefont {Barotta}, \citenamefont {Spradlin},
  \citenamefont {Pucci}, \citenamefont {Hunt},\ and\ \citenamefont
  {Harris}}]{Hooshanginejad24}%
  \BibitemOpen
  \bibfield  {author} {\bibinfo {author} {\bibfnamefont {A.}~\bibnamefont
  {Hooshanginejad}}, \bibinfo {author} {\bibfnamefont {J.-W.}\ \bibnamefont
  {Barotta}}, \bibinfo {author} {\bibfnamefont {V.}~\bibnamefont {Spradlin}},
  \bibinfo {author} {\bibfnamefont {G.}~\bibnamefont {Pucci}}, \bibinfo
  {author} {\bibfnamefont {R.}~\bibnamefont {Hunt}},\ and\ \bibinfo {author}
  {\bibfnamefont {D.~M.}\ \bibnamefont {Harris}},\ }\bibfield  {title}
  {\bibinfo {title} {Interactions and pattern formation in a macroscopic
  magnetocapillary salr system of mermaid cereal},\ }\href
  {https://doi.org/10.1038/s41467-024-49754-4} {\bibfield  {journal} {\bibinfo
  {journal} {Nature Commun.}\ }\textbf {\bibinfo {volume} {15}},\ \bibinfo
  {pages} {5466} (\bibinfo {year} {2024})}\BibitemShut {NoStop}%
\bibitem [{\citenamefont {Jagla}(1998)}]{Jagla98}%
  \BibitemOpen
  \bibfield  {author} {\bibinfo {author} {\bibfnamefont {E.~A.}\ \bibnamefont
  {Jagla}},\ }\bibfield  {title} {\bibinfo {title} {Phase behavior of a system
  of particles with core collapse},\ }\href
  {https://doi.org/10.1103/PhysRevE.58.1478} {\bibfield  {journal} {\bibinfo
  {journal} {Phys. Rev. E}\ }\textbf {\bibinfo {volume} {58}},\ \bibinfo
  {pages} {1478} (\bibinfo {year} {1998})}\BibitemShut {NoStop}%
\bibitem [{\citenamefont {Malescio}\ and\ \citenamefont
  {Pellicane}(2003)}]{Malescio03}%
  \BibitemOpen
  \bibfield  {author} {\bibinfo {author} {\bibfnamefont {G.}~\bibnamefont
  {Malescio}}\ and\ \bibinfo {author} {\bibfnamefont {G.}~\bibnamefont
  {Pellicane}},\ }\bibfield  {title} {\bibinfo {title} {Stripe phases from
  isotropic repulsive interactions},\ }\href {https://doi.org/10.1038/nmat820}
  {\bibfield  {journal} {\bibinfo  {journal} {Nature Mater.}\ }\textbf
  {\bibinfo {volume} {2}},\ \bibinfo {pages} {97} (\bibinfo {year}
  {2003})}\BibitemShut {NoStop}%
\bibitem [{\citenamefont {Glaser}\ \emph {et~al.}(2007)\citenamefont {Glaser},
  \citenamefont {Grason}, \citenamefont {Kamien}, \citenamefont {Kosmrlj},
  \citenamefont {Santangelo},\ and\ \citenamefont {Ziherl}}]{Glaser07}%
  \BibitemOpen
  \bibfield  {author} {\bibinfo {author} {\bibfnamefont {M.~A.}\ \bibnamefont
  {Glaser}}, \bibinfo {author} {\bibfnamefont {G.~M.}\ \bibnamefont {Grason}},
  \bibinfo {author} {\bibfnamefont {R.~D.}\ \bibnamefont {Kamien}}, \bibinfo
  {author} {\bibfnamefont {A.}~\bibnamefont {Kosmrlj}}, \bibinfo {author}
  {\bibfnamefont {C.~D.}\ \bibnamefont {Santangelo}},\ and\ \bibinfo {author}
  {\bibfnamefont {P.}~\bibnamefont {Ziherl}},\ }\bibfield  {title} {\bibinfo
  {title} {Soft spheres make more mesophases},\ }\href
  {https://doi.org/10.1209/0295-5075/78/46004} {\bibfield  {journal} {\bibinfo
  {journal} {EPL}\ }\textbf {\bibinfo {volume} {78}},\ \bibinfo {pages} {46004}
  (\bibinfo {year} {2007})}\BibitemShut {NoStop}%
\bibitem [{\citenamefont {Costa~Campos}\ \emph {et~al.}(2013)\citenamefont
  {Costa~Campos}, \citenamefont {Apolinario},\ and\ \citenamefont
  {L\"owen}}]{CostaCampos13}%
  \BibitemOpen
  \bibfield  {author} {\bibinfo {author} {\bibfnamefont {L.~Q.}\ \bibnamefont
  {Costa~Campos}}, \bibinfo {author} {\bibfnamefont {S.~W.~S.}\ \bibnamefont
  {Apolinario}},\ and\ \bibinfo {author} {\bibfnamefont {H.}~\bibnamefont
  {L\"owen}},\ }\bibfield  {title} {\bibinfo {title} {Structural ordering of
  trapped colloids with competing interactions},\ }\href
  {https://doi.org/10.1103/PhysRevE.88.042313} {\bibfield  {journal} {\bibinfo
  {journal} {Phys. Rev. E}\ }\textbf {\bibinfo {volume} {88}},\ \bibinfo
  {pages} {042313} (\bibinfo {year} {2013})}\BibitemShut {NoStop}%
\bibitem [{\citenamefont {Fogler}\ \emph {et~al.}(1996)\citenamefont {Fogler},
  \citenamefont {Koulakov},\ and\ \citenamefont {Shklovskii}}]{Fogler96}%
  \BibitemOpen
  \bibfield  {author} {\bibinfo {author} {\bibfnamefont {M.~M.}\ \bibnamefont
  {Fogler}}, \bibinfo {author} {\bibfnamefont {A.~A.}\ \bibnamefont
  {Koulakov}},\ and\ \bibinfo {author} {\bibfnamefont {B.~I.}\ \bibnamefont
  {Shklovskii}},\ }\bibfield  {title} {\bibinfo {title} {Ground state of a
  two-dimensional electron liquid in a weak magnetic field},\ }\href
  {https://doi.org/10.1103/PhysRevB.54.1853} {\bibfield  {journal} {\bibinfo
  {journal} {Phys. Rev. B}\ }\textbf {\bibinfo {volume} {54}},\ \bibinfo
  {pages} {1853} (\bibinfo {year} {1996})}\BibitemShut {NoStop}%
\bibitem [{\citenamefont {Moessner}\ and\ \citenamefont
  {Chalker}(1996)}]{Moessner96}%
  \BibitemOpen
  \bibfield  {author} {\bibinfo {author} {\bibfnamefont {R.}~\bibnamefont
  {Moessner}}\ and\ \bibinfo {author} {\bibfnamefont {J.~T.}\ \bibnamefont
  {Chalker}},\ }\bibfield  {title} {\bibinfo {title} {Exact results for
  interacting electrons in high {L}andau levels},\ }\href
  {https://doi.org/10.1103/PhysRevB.54.5006} {\bibfield  {journal} {\bibinfo
  {journal} {Phys. Rev. B}\ }\textbf {\bibinfo {volume} {54}},\ \bibinfo
  {pages} {5006} (\bibinfo {year} {1996})}\BibitemShut {NoStop}%
\bibitem [{\citenamefont {Cooper}\ \emph {et~al.}(1999)\citenamefont {Cooper},
  \citenamefont {Lilly}, \citenamefont {Eisenstein}, \citenamefont {Pfeiffer},\
  and\ \citenamefont {West}}]{Cooper99}%
  \BibitemOpen
  \bibfield  {author} {\bibinfo {author} {\bibfnamefont {K.~B.}\ \bibnamefont
  {Cooper}}, \bibinfo {author} {\bibfnamefont {M.~P.}\ \bibnamefont {Lilly}},
  \bibinfo {author} {\bibfnamefont {J.~P.}\ \bibnamefont {Eisenstein}},
  \bibinfo {author} {\bibfnamefont {L.~N.}\ \bibnamefont {Pfeiffer}},\ and\
  \bibinfo {author} {\bibfnamefont {K.~W.}\ \bibnamefont {West}},\ }\bibfield
  {title} {\bibinfo {title} {Insulating phases of two-dimensional electrons in
  high {L}andau levels: Observation of sharp thresholds to conduction},\ }\href
  {https://doi.org/10.1103/PhysRevB.60.R11285} {\bibfield  {journal} {\bibinfo
  {journal} {Phys. Rev. B}\ }\textbf {\bibinfo {volume} {60}},\ \bibinfo
  {pages} {R11285} (\bibinfo {year} {1999})}\BibitemShut {NoStop}%
\bibitem [{\citenamefont {Fradkin}\ and\ \citenamefont
  {Kivelson}(1999)}]{Fradkin99}%
  \BibitemOpen
  \bibfield  {author} {\bibinfo {author} {\bibfnamefont {E.}~\bibnamefont
  {Fradkin}}\ and\ \bibinfo {author} {\bibfnamefont {S.~A.}\ \bibnamefont
  {Kivelson}},\ }\bibfield  {title} {\bibinfo {title} {Liquid-crystal phases of
  quantum {Hall} systems},\ }\href {https://doi.org/10.1103/PhysRevB.59.8065}
  {\bibfield  {journal} {\bibinfo  {journal} {Phys. Rev. B}\ }\textbf {\bibinfo
  {volume} {59}},\ \bibinfo {pages} {8065} (\bibinfo {year}
  {1999})}\BibitemShut {NoStop}%
\bibitem [{\citenamefont {G\"ores}\ \emph {et~al.}(2007)\citenamefont
  {G\"ores}, \citenamefont {Gamez}, \citenamefont {Smet}, \citenamefont
  {Pfeiffer}, \citenamefont {West}, \citenamefont {Yacoby}, \citenamefont
  {Umansky},\ and\ \citenamefont {von Klitzing}}]{Gores07}%
  \BibitemOpen
  \bibfield  {author} {\bibinfo {author} {\bibfnamefont {J.}~\bibnamefont
  {G\"ores}}, \bibinfo {author} {\bibfnamefont {G.}~\bibnamefont {Gamez}},
  \bibinfo {author} {\bibfnamefont {J.~H.}\ \bibnamefont {Smet}}, \bibinfo
  {author} {\bibfnamefont {L.}~\bibnamefont {Pfeiffer}}, \bibinfo {author}
  {\bibfnamefont {K.}~\bibnamefont {West}}, \bibinfo {author} {\bibfnamefont
  {A.}~\bibnamefont {Yacoby}}, \bibinfo {author} {\bibfnamefont
  {V.}~\bibnamefont {Umansky}},\ and\ \bibinfo {author} {\bibfnamefont
  {K.}~\bibnamefont {von Klitzing}},\ }\bibfield  {title} {\bibinfo {title}
  {Current-induced anisotropy and reordering of the electron liquid-crystal
  phases in a two-dimensional electron system},\ }\href
  {https://doi.org/10.1103/PhysRevLett.99.246402} {\bibfield  {journal}
  {\bibinfo  {journal} {Phys. Rev. Lett.}\ }\textbf {\bibinfo {volume} {99}},\
  \bibinfo {pages} {246402} (\bibinfo {year} {2007})}\BibitemShut {NoStop}%
\bibitem [{\citenamefont {Zhu}\ \emph {et~al.}(2009)\citenamefont {Zhu},
  \citenamefont {Sambandamurthy}, \citenamefont {Engel}, \citenamefont {Tsui},
  \citenamefont {Pfeiffer},\ and\ \citenamefont {West}}]{Zhu09}%
  \BibitemOpen
  \bibfield  {author} {\bibinfo {author} {\bibfnamefont {H.}~\bibnamefont
  {Zhu}}, \bibinfo {author} {\bibfnamefont {G.}~\bibnamefont {Sambandamurthy}},
  \bibinfo {author} {\bibfnamefont {L.~W.}\ \bibnamefont {Engel}}, \bibinfo
  {author} {\bibfnamefont {D.~C.}\ \bibnamefont {Tsui}}, \bibinfo {author}
  {\bibfnamefont {L.~N.}\ \bibnamefont {Pfeiffer}},\ and\ \bibinfo {author}
  {\bibfnamefont {K.~W.}\ \bibnamefont {West}},\ }\bibfield  {title} {\bibinfo
  {title} {Pinning mode resonances of {2D} electron stripe phases: Effect of an
  in-plane magnetic field},\ }\href
  {https://doi.org/10.1103/PhysRevLett.102.136804} {\bibfield  {journal}
  {\bibinfo  {journal} {Phys. Rev. Lett.}\ }\textbf {\bibinfo {volume} {102}},\
  \bibinfo {pages} {136804} (\bibinfo {year} {2009})}\BibitemShut {NoStop}%
\bibitem [{\citenamefont {Friess}\ \emph {et~al.}(2018)\citenamefont {Friess},
  \citenamefont {Umansky}, \citenamefont {von Klitzing},\ and\ \citenamefont
  {Smet}}]{Friess18}%
  \BibitemOpen
  \bibfield  {author} {\bibinfo {author} {\bibfnamefont {B.}~\bibnamefont
  {Friess}}, \bibinfo {author} {\bibfnamefont {V.}~\bibnamefont {Umansky}},
  \bibinfo {author} {\bibfnamefont {K.}~\bibnamefont {von Klitzing}},\ and\
  \bibinfo {author} {\bibfnamefont {J.~H.}\ \bibnamefont {Smet}},\ }\bibfield
  {title} {\bibinfo {title} {Current flow in the bubble and stripe phases},\
  }\href {https://doi.org/10.1103/PhysRevLett.120.137603} {\bibfield  {journal}
  {\bibinfo  {journal} {Phys. Rev. Lett.}\ }\textbf {\bibinfo {volume} {120}},\
  \bibinfo {pages} {137603} (\bibinfo {year} {2018})}\BibitemShut {NoStop}%
\bibitem [{\citenamefont {Tranquada}\ \emph {et~al.}(1995)\citenamefont
  {Tranquada}, \citenamefont {Sterlieb}, \citenamefont {Axe}, \citenamefont
  {Nakamura},\ and\ \citenamefont {Uchida}}]{Tranquada95}%
  \BibitemOpen
  \bibfield  {author} {\bibinfo {author} {\bibfnamefont {J.~M.}\ \bibnamefont
  {Tranquada}}, \bibinfo {author} {\bibfnamefont {B.~J.}\ \bibnamefont
  {Sterlieb}}, \bibinfo {author} {\bibfnamefont {J.~D.}\ \bibnamefont {Axe}},
  \bibinfo {author} {\bibfnamefont {Y.}~\bibnamefont {Nakamura}},\ and\
  \bibinfo {author} {\bibfnamefont {S.}~\bibnamefont {Uchida}},\ }\bibfield
  {title} {\bibinfo {title} {Evidence for stripe correlations of spins and
  holes in copper-oxide superconductors},\ }\href
  {https://doi.org/10.1038/375561a0} {\bibfield  {journal} {\bibinfo  {journal}
  {Nature (London)}\ }\textbf {\bibinfo {volume} {375}},\ \bibinfo {pages}
  {561} (\bibinfo {year} {1995})}\BibitemShut {NoStop}%
\bibitem [{\citenamefont {Olson~Reichhardt}\ \emph {et~al.}(2004)\citenamefont
  {Olson~Reichhardt}, \citenamefont {Reichhardt},\ and\ \citenamefont
  {Bishop}}]{Reichhardt04a}%
  \BibitemOpen
  \bibfield  {author} {\bibinfo {author} {\bibfnamefont {C.~J.}\ \bibnamefont
  {Olson~Reichhardt}}, \bibinfo {author} {\bibfnamefont {C.}~\bibnamefont
  {Reichhardt}},\ and\ \bibinfo {author} {\bibfnamefont {A.~R.}\ \bibnamefont
  {Bishop}},\ }\bibfield  {title} {\bibinfo {title} {Fibrillar templates and
  soft phases in systems with short-range dipolar and long-range
  interactions},\ }\href {https://doi.org/10.1103/PhysRevLett.92.016801}
  {\bibfield  {journal} {\bibinfo  {journal} {Phys. Rev. Lett.}\ }\textbf
  {\bibinfo {volume} {92}},\ \bibinfo {pages} {016801} (\bibinfo {year}
  {2004})}\BibitemShut {NoStop}%
\bibitem [{\citenamefont {Mertelj}\ \emph {et~al.}(2005)\citenamefont
  {Mertelj}, \citenamefont {Kabanov},\ and\ \citenamefont
  {Mihailovic}}]{Mertelj05}%
  \BibitemOpen
  \bibfield  {author} {\bibinfo {author} {\bibfnamefont {T.}~\bibnamefont
  {Mertelj}}, \bibinfo {author} {\bibfnamefont {V.~V.}\ \bibnamefont
  {Kabanov}},\ and\ \bibinfo {author} {\bibfnamefont {D.}~\bibnamefont
  {Mihailovic}},\ }\bibfield  {title} {\bibinfo {title} {Charged particles on a
  two-dimensional lattice subject to anisotropic {Jahn-T}eller interactions},\
  }\href {https://doi.org/10.1103/PhysRevLett.94.147003} {\bibfield  {journal}
  {\bibinfo  {journal} {Phys. Rev. Lett.}\ }\textbf {\bibinfo {volume} {94}},\
  \bibinfo {pages} {147003} (\bibinfo {year} {2005})}\BibitemShut {NoStop}%
\bibitem [{\citenamefont {Xu}\ \emph {et~al.}(2011)\citenamefont {Xu},
  \citenamefont {Fangohr}, \citenamefont {Ding}, \citenamefont {Zhou},
  \citenamefont {Xu}, \citenamefont {Wang}, \citenamefont {Gu}, \citenamefont
  {Shi},\ and\ \citenamefont {Dou}}]{Xu11}%
  \BibitemOpen
  \bibfield  {author} {\bibinfo {author} {\bibfnamefont {X.~B.}\ \bibnamefont
  {Xu}}, \bibinfo {author} {\bibfnamefont {H.}~\bibnamefont {Fangohr}},
  \bibinfo {author} {\bibfnamefont {S.~Y.}\ \bibnamefont {Ding}}, \bibinfo
  {author} {\bibfnamefont {F.}~\bibnamefont {Zhou}}, \bibinfo {author}
  {\bibfnamefont {X.~N.}\ \bibnamefont {Xu}}, \bibinfo {author} {\bibfnamefont
  {Z.~H.}\ \bibnamefont {Wang}}, \bibinfo {author} {\bibfnamefont
  {M.}~\bibnamefont {Gu}}, \bibinfo {author} {\bibfnamefont {D.~Q.}\
  \bibnamefont {Shi}},\ and\ \bibinfo {author} {\bibfnamefont {S.~X.}\
  \bibnamefont {Dou}},\ }\bibfield  {title} {\bibinfo {title} {Phase diagram of
  vortex matter of type-{II} superconductors},\ }\href
  {https://doi.org/10.1103/PhysRevB.83.014501} {\bibfield  {journal} {\bibinfo
  {journal} {Phys. Rev. B}\ }\textbf {\bibinfo {volume} {83}},\ \bibinfo
  {pages} {014501} (\bibinfo {year} {2011})}\BibitemShut {NoStop}%
\bibitem [{\citenamefont {Komendov\'a}\ \emph {et~al.}(2013)\citenamefont
  {Komendov\'a}, \citenamefont {Milo\ifmmode \check{s}\else
  \v{s}\fi{}evi\ifmmode~\acute{c}\else \'{c}\fi{}},\ and\ \citenamefont
  {Peeters}}]{Komendova13}%
  \BibitemOpen
  \bibfield  {author} {\bibinfo {author} {\bibfnamefont {L.}~\bibnamefont
  {Komendov\'a}}, \bibinfo {author} {\bibfnamefont {M.~V.}\ \bibnamefont
  {Milo\ifmmode \check{s}\else \v{s}\fi{}evi\ifmmode~\acute{c}\else
  \'{c}\fi{}}},\ and\ \bibinfo {author} {\bibfnamefont {F.~M.}\ \bibnamefont
  {Peeters}},\ }\bibfield  {title} {\bibinfo {title} {Soft vortex matter in a
  type-{I}/type-{II} superconducting bilayer},\ }\href
  {https://doi.org/10.1103/PhysRevB.88.094515} {\bibfield  {journal} {\bibinfo
  {journal} {Phys. Rev. B}\ }\textbf {\bibinfo {volume} {88}},\ \bibinfo
  {pages} {094515} (\bibinfo {year} {2013})}\BibitemShut {NoStop}%
\bibitem [{\citenamefont {Varney}\ \emph {et~al.}(2013)\citenamefont {Varney},
  \citenamefont {Sellin}, \citenamefont {Wang}, \citenamefont {Fangohr},\ and\
  \citenamefont {Babaev}}]{Varney13}%
  \BibitemOpen
  \bibfield  {author} {\bibinfo {author} {\bibfnamefont {C.~N.}\ \bibnamefont
  {Varney}}, \bibinfo {author} {\bibfnamefont {K.~A.~H.}\ \bibnamefont
  {Sellin}}, \bibinfo {author} {\bibfnamefont {Q.-Z.}\ \bibnamefont {Wang}},
  \bibinfo {author} {\bibfnamefont {H.}~\bibnamefont {Fangohr}},\ and\ \bibinfo
  {author} {\bibfnamefont {E.}~\bibnamefont {Babaev}},\ }\bibfield  {title}
  {\bibinfo {title} {Hierarchical structure foramtion in layered
  superconducting systems with multi-scale inter-vortex interactions},\ }\href
  {https://doi.org/10.1088/0953-8984/25/41/415702} {\bibfield  {journal}
  {\bibinfo  {journal} {J. Phys.: Condens. Matter}\ }\textbf {\bibinfo {volume}
  {25}},\ \bibinfo {pages} {415702} (\bibinfo {year} {2013})}\BibitemShut
  {NoStop}%
\bibitem [{\citenamefont {Sellin}\ and\ \citenamefont
  {Babaev}(2013)}]{Sellin13}%
  \BibitemOpen
  \bibfield  {author} {\bibinfo {author} {\bibfnamefont {K.~A.~H.}\
  \bibnamefont {Sellin}}\ and\ \bibinfo {author} {\bibfnamefont
  {E.}~\bibnamefont {Babaev}},\ }\bibfield  {title} {\bibinfo {title} {Stripe,
  gossamer, and glassy phases in systems with strong nonpairwise
  interactions},\ }\href {https://doi.org/10.1103/PhysRevE.88.042305}
  {\bibfield  {journal} {\bibinfo  {journal} {Phys. Rev. E}\ }\textbf {\bibinfo
  {volume} {88}},\ \bibinfo {pages} {042305} (\bibinfo {year}
  {2013})}\BibitemShut {NoStop}%
\bibitem [{\citenamefont {Brems}\ \emph {et~al.}(2022)\citenamefont {Brems},
  \citenamefont {M{\" u}hlbauer}, \citenamefont {C{\' o}rdoba-Camacho},
  \citenamefont {Shanenko}, \citenamefont {Vagov}, \citenamefont {Aguiar},\
  and\ \citenamefont {Cubitt}}]{Brems22}%
  \BibitemOpen
  \bibfield  {author} {\bibinfo {author} {\bibfnamefont {X.~S.}\ \bibnamefont
  {Brems}}, \bibinfo {author} {\bibfnamefont {S.}~\bibnamefont {M{\"
  u}hlbauer}}, \bibinfo {author} {\bibfnamefont {W.~Y.}\ \bibnamefont {C{\'
  o}rdoba-Camacho}}, \bibinfo {author} {\bibfnamefont {A.~A.}\ \bibnamefont
  {Shanenko}}, \bibinfo {author} {\bibfnamefont {A.}~\bibnamefont {Vagov}},
  \bibinfo {author} {\bibfnamefont {J.~A.}\ \bibnamefont {Aguiar}},\ and\
  \bibinfo {author} {\bibfnamefont {R.}~\bibnamefont {Cubitt}},\ }\bibfield
  {title} {\bibinfo {title} {Current-induced self-organisation of mixed
  superconducting states},\ }\href {https://doi.org/10.1088/1361-6668/ac455e}
  {\bibfield  {journal} {\bibinfo  {journal} {Supercond. Sci. Technol.}\
  }\textbf {\bibinfo {volume} {35}},\ \bibinfo {pages} {035003} (\bibinfo
  {year} {2022})}\BibitemShut {NoStop}%
\bibitem [{\citenamefont {Reichhardt}\ \emph {et~al.}(2022)\citenamefont
  {Reichhardt}, \citenamefont {Reichhardt},\ and\ \citenamefont
  {Milosevic}}]{Reichhardt22a}%
  \BibitemOpen
  \bibfield  {author} {\bibinfo {author} {\bibfnamefont {C.}~\bibnamefont
  {Reichhardt}}, \bibinfo {author} {\bibfnamefont {C.~J.~O.}\ \bibnamefont
  {Reichhardt}},\ and\ \bibinfo {author} {\bibfnamefont {M.}~\bibnamefont
  {Milosevic}},\ }\bibfield  {title} {\bibinfo {title} {Statics and dynamics of
  skyrmions interacting with disorder and nanostructures},\ }\href
  {https://doi.org/10.1103/RevModPhys.94.035005} {\bibfield  {journal}
  {\bibinfo  {journal} {Rev. Mod. Phys.}\ }\textbf {\bibinfo {volume} {94}},\
  \bibinfo {pages} {035005} (\bibinfo {year} {2022})}\BibitemShut {NoStop}%
\bibitem [{\citenamefont {McDermott}\ \emph {et~al.}(2016)\citenamefont
  {McDermott}, \citenamefont {Reichhardt},\ and\ \citenamefont
  {Reichhardt}}]{McDermott16}%
  \BibitemOpen
  \bibfield  {author} {\bibinfo {author} {\bibfnamefont {D.}~\bibnamefont
  {McDermott}}, \bibinfo {author} {\bibfnamefont {C.~J.~O.}\ \bibnamefont
  {Reichhardt}},\ and\ \bibinfo {author} {\bibfnamefont {C.}~\bibnamefont
  {Reichhardt}},\ }\bibfield  {title} {\bibinfo {title} {Structural transitions
  and hysteresis in clump- and stripe-forming systems under dynamic
  compression},\ }\href {https://doi.org/10.1039/C6SM01939K} {\bibfield
  {journal} {\bibinfo  {journal} {Soft Matter}\ }\textbf {\bibinfo {volume}
  {12}},\ \bibinfo {pages} {9549} (\bibinfo {year} {2016})}\BibitemShut
  {NoStop}%
\bibitem [{\citenamefont {Reichhardt}\ and\ \citenamefont
  {Reichhardt}(2024)}]{Reichhardt24}%
  \BibitemOpen
  \bibfield  {author} {\bibinfo {author} {\bibfnamefont {C.}~\bibnamefont
  {Reichhardt}}\ and\ \bibinfo {author} {\bibfnamefont {C.~J.~O.}\ \bibnamefont
  {Reichhardt}},\ }\bibfield  {title} {\bibinfo {title} {Peak effect and
  dynamics of stripe- and pattern-forming systems on a periodic one-dimensional
  substrate},\ }\href {https://doi.org/10.1103/PhysRevE.109.054606} {\bibfield
  {journal} {\bibinfo  {journal} {Phys. Rev. E}\ }\textbf {\bibinfo {volume}
  {109}},\ \bibinfo {pages} {054606} (\bibinfo {year} {2024})}\BibitemShut
  {NoStop}%
\bibitem [{\citenamefont {Reichhardt}\ \emph
  {et~al.}(2003{\natexlab{b}})\citenamefont {Reichhardt}, \citenamefont
  {Reichhardt}, \citenamefont {Martin},\ and\ \citenamefont
  {Bishop}}]{Reichhardt03a}%
  \BibitemOpen
  \bibfield  {author} {\bibinfo {author} {\bibfnamefont {C.}~\bibnamefont
  {Reichhardt}}, \bibinfo {author} {\bibfnamefont {C.~J.~O.}\ \bibnamefont
  {Reichhardt}}, \bibinfo {author} {\bibfnamefont {I.}~\bibnamefont {Martin}},\
  and\ \bibinfo {author} {\bibfnamefont {A.~R.}\ \bibnamefont {Bishop}},\
  }\bibfield  {title} {\bibinfo {title} {Dynamical ordering of driven stripe
  phases in quenched disorder},\ }\href
  {https://doi.org/10.1103/PhysRevLett.90.026401} {\bibfield  {journal}
  {\bibinfo  {journal} {Phys. Rev. Lett.}\ }\textbf {\bibinfo {volume} {90}},\
  \bibinfo {pages} {026401} (\bibinfo {year} {2003}{\natexlab{b}})}\BibitemShut
  {NoStop}%
\end{thebibliography}%

\end{document}